\documentclass{aa}

\usepackage{graphicx}
\usepackage{txfonts}
\usepackage{natbib}

\begin{document}
\title{Gas temperature profiles in galaxy clusters with Swift XRT: observations and capabilities 
to map near R$_{\rm200}$ }

%\thanks{}
\author{
A. Moretti\inst{1}, F. Gastaldello\inst{2}, S. Ettori\inst{3,4}, S. Molendi \inst{2}} 

\offprints{alberto.moretti@brera.inaf.it}

\institute{
INAF, Osservatorio Astronomico di Brera, Via E. Bianchi 46, 23807, Merate (LC), Italy
\and
INAF, Via E. Bassini 15, 20133 Milano, Italy
\and
INAF, Osservatorio Astronomico di Bologna, Via Ranzani 1, 40127, Bologna, Italy
\and
INFN, Sezione di Bologna, viale Berti Pichat 6/2, I-40127 Bologna, Italy
%Department of Astronomy \& Astrophysics, Pennsylvania State University, 525 Davey Lab, University Park, PA 16802, USA
%\and
%INAF, Istituto di Astrofisica Spaziale e Fisica Cosmica Sezione di Palermo, Via U.\ La Malfa 153, I-90146 Palermo, Italy
%\and
%ASI Science Data Center, via G.\ Galilei, I-00044 Frascati, Italy
%\and  
%University of Leicester, LE1 7RH, UK
%\and
%Stanford Linear Accelerator Center, 2575 Sand Hill Road, Menlo Park, CA 94025
%\and
%KIPAC, 2575 Sand Hill Road, Menlo Park, CA 94025
%\and 
%Universit\`a degli Studi di Milano-Bicocca, Dipartimento di Fisica, Piazza delle Scienze 3, I-20126 Milano, Italy
%\and 
%NASA/Goddard Space Flight Center, Greenbelt Road, Greenbelt, MD20771, USA
%\and       
%Department of Physics and Astronomy, Johns Hopkins University, 3400 Charles Street, Baltimore, MD 21218, USA
%\and 
%Universities Space Research Association, 10211 Wincopin Circle, Suite 500, Columbia, MD, 21044-3432, USA
%
%\and
%X-Ray Observational Astronomy Group, Department of Physics \& Astronomy, University of Leicester, LE1 7RH, UK
}
\date{Received ; accepted }
\date{Received ; accepted }
\titlerunning{Galaxy clusters external regions}
\authorrunning{Moretti et al.}  
\abstract{} 
{We investigate the possibility of using the X-ray telescope (XRT) on board the Swift satellite to improve the current accuracy of the intra-cluster 
medium (ICM) temperature measurements in the region close to the virial radius of nearby clusters.}
{We present the spectral analysis of the Swift XRT observations of 6 galaxy clusters and their temperature profiles in the regions within 0.2-0.6 r$_{200}$.
Four of them are nearby famous and very well studied objects (Coma, Abell 1795, Abell 2029 and PKS0745-19). The remaining two, SWJ1557+35 and SWJ0847+13, at redshift 
z=0.16 and z=0.36,  were serendipitously observed by Swift-XRT. 
We accurately quantify the temperature uncertainties, with particular focus on the impact of  the background scatter (both instrumental and cosmic). 
We extrapolate these results and simulate a deep observation of the external region of Abell 1795 which is assumed here as a case study. 
In particular we calculate the expected uncertainties in the temperature measurement as far as  r$_{200}$. }
{We find that, with a fairly deep observation (300 ks), the Swift XRT would be able to measure the ICM temperature profiles in the external regions as 
far as the virial radius, significantly improving the best accuracy among the previous measurements. This can be achieved thanks to the unprecedented 
combination of  good PSF over the full field of view and very accurate control of  the instrumental background. }
{Somehow unexpectedly we conclude that, among currently operating telescope, the Swift-XRT is the only potentially able to improve the current accuracy in plasma temperature measurement at the edges of the  cluster potential. This will be true until a new generation of low-background and large field of view  telescopes, aimed to the study 
of galaxy clusters, will operate. These observations would be of great importance in developing the observing strategy for such missions.}
\maketitle
%%%%%%%%%%%%%%%%%%%%%%%%%%%%%%%%%%%%%%%%%%%%%%%%%%%%%%%%%%%%%%%%%%%%%%%%%%%%%%%%%%%%%%%%%%%%%%%%%%%%%%%%%%%%%%%%%%%%%%%
\section{Introduction}
Galaxy clusters form by the hierarchical accretion of cosmic matter. They reach the virial equilibrium over a volume that defines  the regions where the pristine gas accretes on the dark matter (DM) halo through gravitational collapse and is heated up to millions degrees through adiabatic compression and shocks. 
The end products of this accretion process exhibit in the X-ray band similar radial profiles of surface brightness \citep{Vikhlinin99, Neumann05, Ettori09},
plasma temperature \citep[e.g.]{Allen01, Vikhlinin05, Leccardi09} and gravitational mass distribution \citep[e.g.]{Pointecouteau05}.
\begin{figure*}
\begin{tabular}{ccc}
\includegraphics[width=5.75cm]{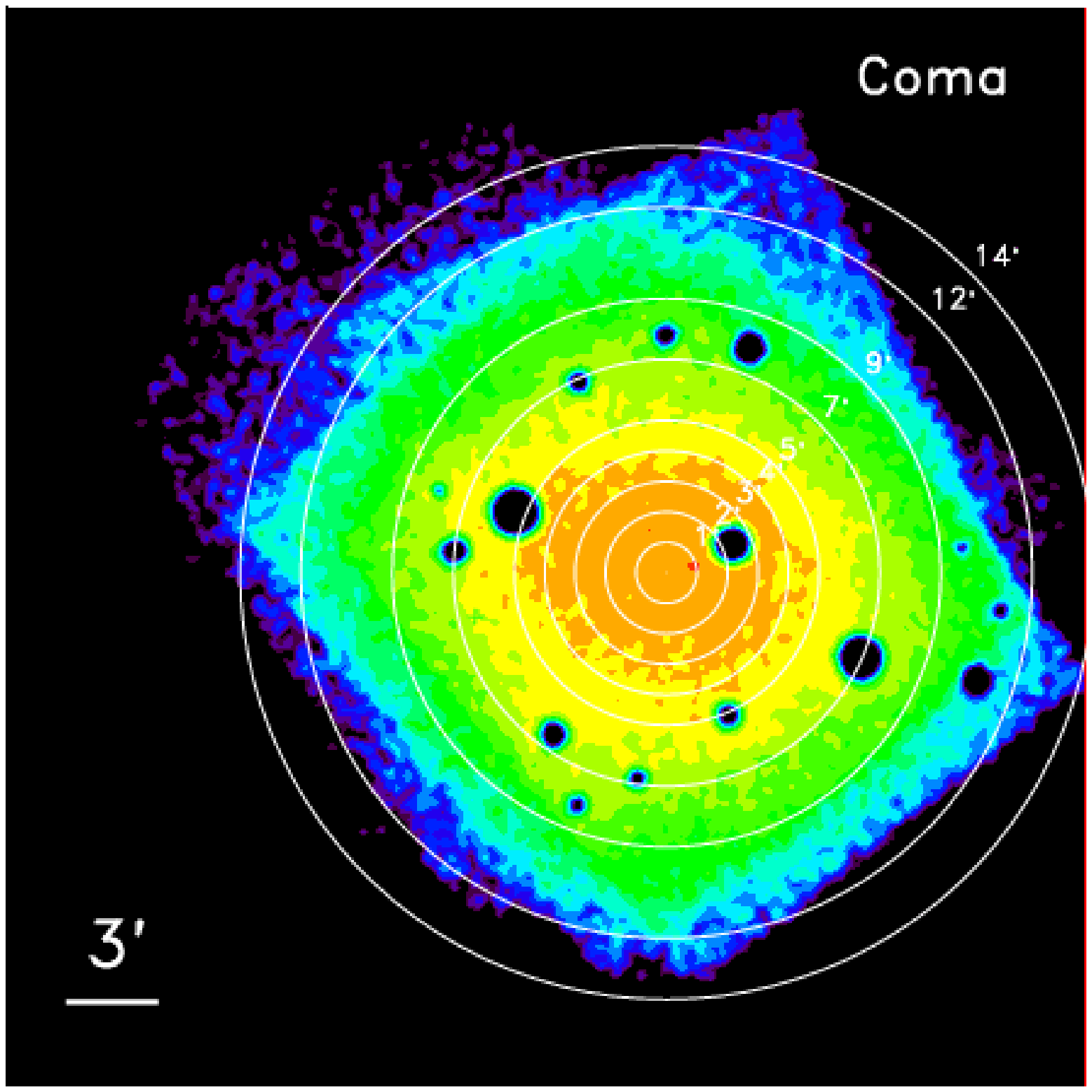}       & \includegraphics[width=5.75cm] {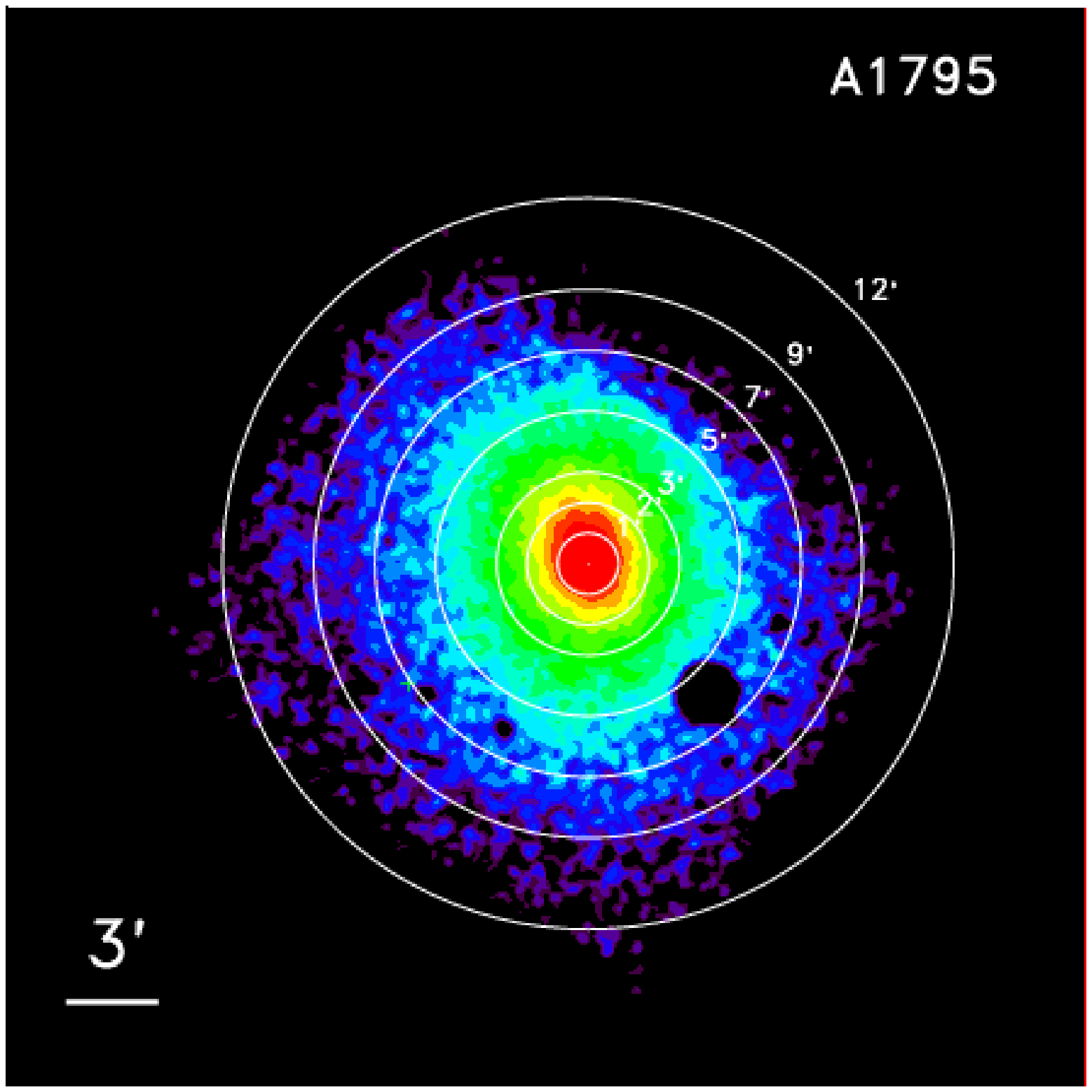}     & \includegraphics[width=5.75cm]{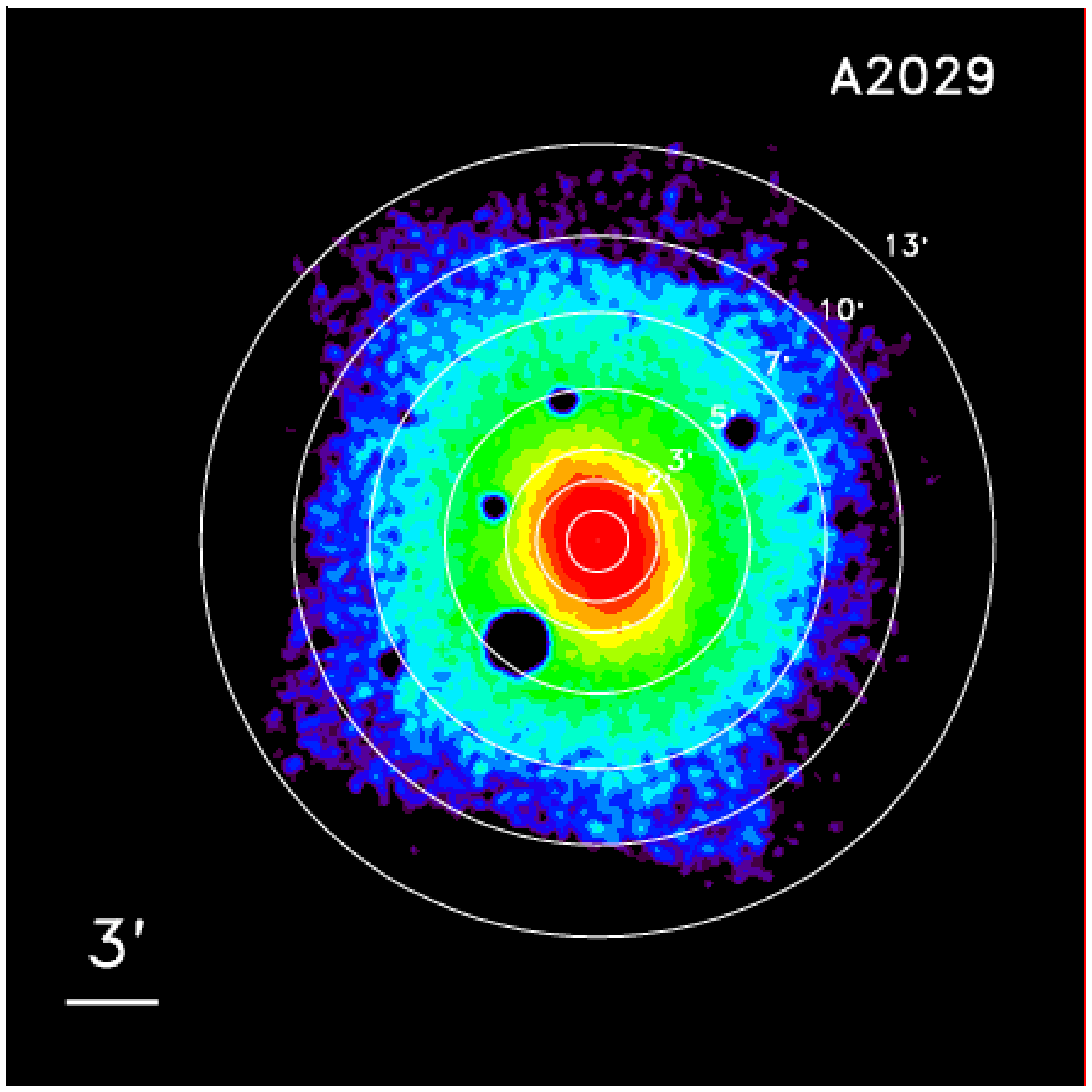}  \\
\includegraphics[width=5.75cm]{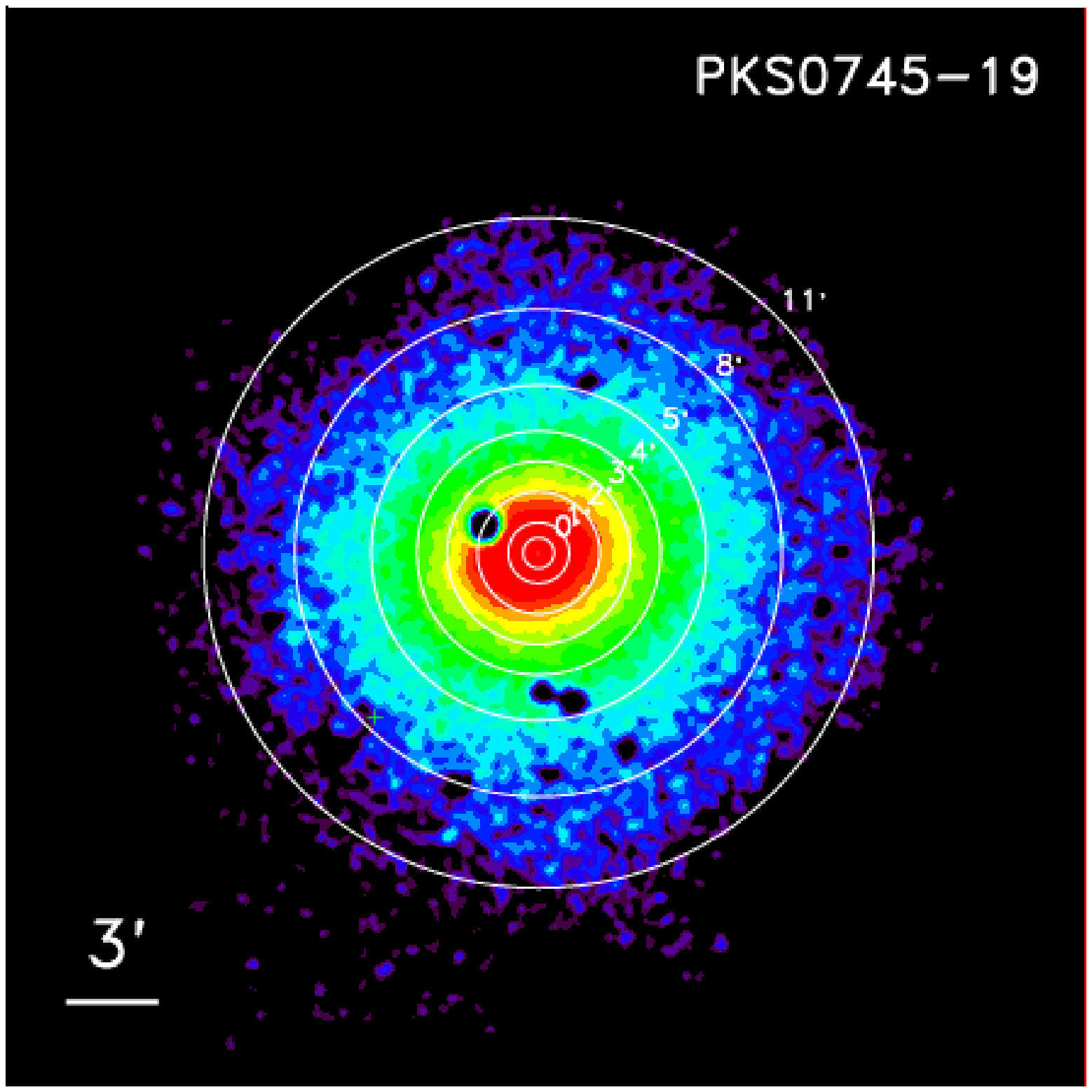}  & \includegraphics[width=5.75cm]{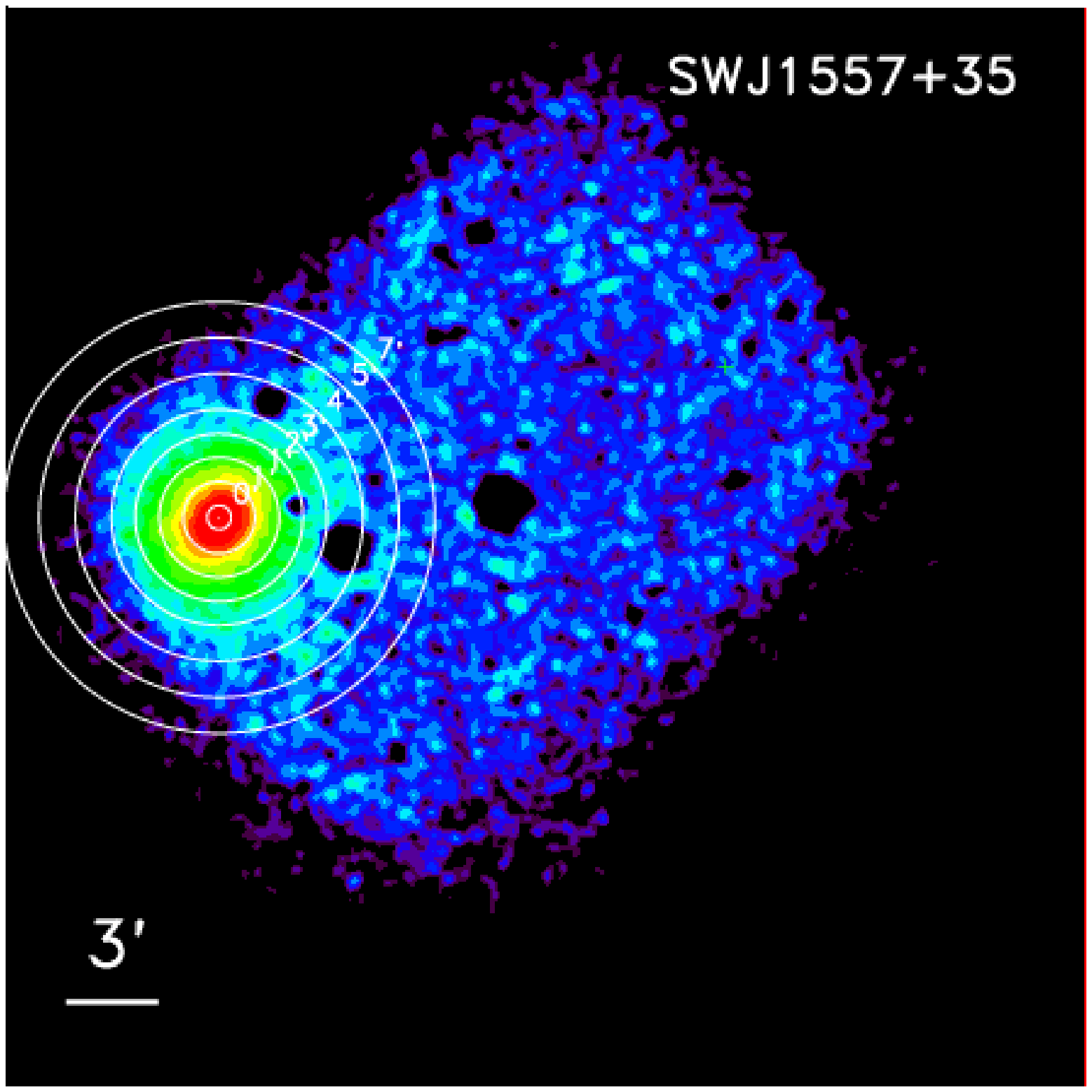}   & \includegraphics[width=5.75cm]{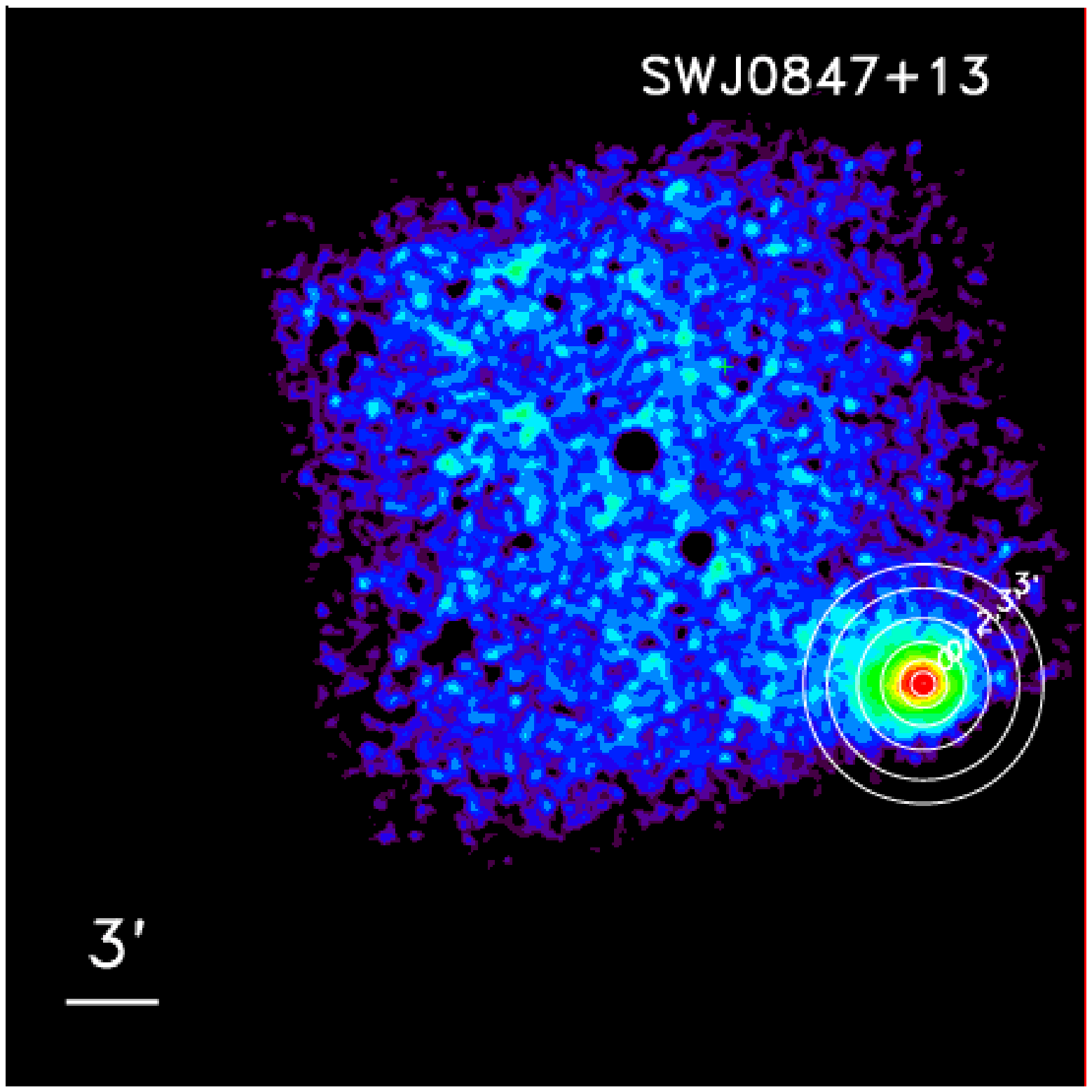}   \\ 
\end{tabular}
\caption{Images for the six objects in our sample, with detected sources excised. Data are smoothed with a 30\arcsec Gaussian filter. 
Superimposed are the circular extraction regions used for the spectral analysis. Circular regions around the detected sources are excised.}
\label{fig:imaprof}
\end{figure*}
\noindent The measurement of the properties of the ICM have been enormously improved thanks to the arcsec resolution and large collecting
area of Chandra and XMM-Newton, but still remain possible only where the X-ray emission can  be well resolved against the background 
(both instrumental and cosmic). While the X-ray surface brightness and gas density can be estimated in few cases above $0.7 r_{200}$ 
\citep[e.g.]{Vikhlinin99, Neumann05, Ettori09}, the ICM temperature, requiring more than an order of magnitude in net counts 
 than the surface brightness to be firmly measured, can be reasonably well constrained up to a fraction ($\sim 0.5-0.6$) of the virial radius 
\citep[e.g.]{Vikhlinin06, Leccardi09, Ettori10, Arnaud10}. The regions not-yet observed are expected to retain most of the information on the processes 
that characterize the accretion and evolution within the cluster of the main baryonic component \citep{Roncarelli06, Rasheed10}.
\begin{table*}
\begin{center}
\caption[]{Cluster sample. 
Mean spectroscopic temperatures and redshift of the first four clusters are from the references reported in the fifth column, through a query to  
BAX archive (\texttt{http://bax.ast.obs-mip.fr/});  the remaining 2 are calculated from our data (see text).
r$_{200}$ are estimated by adopting the scaling relations \citep{Arnaud05} and here are  expressed in units of arcmin and Mpc. 
r$_{max}$ is the center of the most external annulus for which we measured the temperature, here expressed in units of  r$_{200}$. } 
\begin{tabular}{|l|cccccccc|}
\hline
Name              &  RA, Dec                          &   z    &  kT               &   references                &  r$_{200}$         &r$_{max}$ & Exp.& N$_{\rm H}$ \\
                      &    [deg]                            &         & [keV]            &                                  &[$\arcmin$, Mpc]&[r$_{200}]$& [ks] &10$^{20}$cm$^{-2}$\\ 
\hline
Coma              &194.9392 +27.9429&0.023                  & 8.25$\pm$0.01                 &\cite{Arnaud01}  & 78.2 [2.18]   &0.17  & 43.7   & 0.80\\
Abell 1795       &207.2183 +26.5903&0.062                  & 6.12$\pm$0.05                 &\cite{Vikhlinin06}& 25.7 [1.84]   &0.41  &  20.3 & 1.32\\
Abell 2029       &227.7336 +5.74440&0.077                  & 8.47$\pm$0.09                 &\cite{Vikhlinin06}& 26.7 [2.16]   &0.43  & 42.3  & 3.25 \\           
PKS0745-19     &116.8799  -19.2948&0.102                  & 7.97$\pm$0.28                 &\cite{Arnaud05}  & 18.1 [2.06]   &0.58  & 63.1  & 41.8\\
SWJ1557+3530 &239.4287 +35.5073&0.153$\pm0.006$& 6.79$\pm$0.25& this work& 11.5 [1.89]   &0.62  & 180.1 & 0.20\\
SWJ0847+1331 &131.9550 +13.5278&0.358$\pm0.005$& 6.02$\pm$0.34& this work&  5.3  [1.56]   &0.50  & 203.1 & 3.23\\
\hline
\end{tabular}
\label{tab:obse} 
\end{center}
\end{table*}
It is therefore crucial to obtain direct measurements of the cluster properties at these large radii where very important 
processes for the evolution of the clusters take place. Very recently, Suzaku, thanks to its low background and high sensitivity, 
has been able to map roughly (i.e. with a spatial resolution limited to $>$4$\arcmin$) the regions close to the virial radius, 
providing the first estimate of the gas temperature in 6 objects \citep{Fujita08, George09, Reiprich09, Bautz09, Kawaharada10, Hoshino10}.
The aim of this paper is to show that the X-ray telescope  (XRT) \citep{Burrows05} on board the  Swift satellite \citep{Gehrels04}  
can improve the accuracy of these measurements with fairly deep observations. 
To this end,  we present the first spectral analysis of the archived Swift observations of nearby clusters (Section~\ref{sect:data}). 
Since this is a non-standard analysis for the Swift - XRT data and is used and presented here for the first time,  we discuss 
in detail the techniques adopted. 
In particular in Section~\ref{sect:backg} we describe the procedure we developed to estimate the background and its systematic uncertainty. 
To calculate how much this uncertainty affects the temperature measurement we performed a series of Monte Carlo simulations of thermal  
spectrum at different level of surface brightness (described in Section~\ref{sect:ssist} ).
Finally in Section~\ref{sect:tsimu} we used these results to properly and robustly  evaluate the expected uncertainties on the gas temperature 
measurements at r$_{\rm200}$
\footnote{The radius that defines the sphere enclosing a mean cluster density that is 200 times the critical value at the cluster's redshift.
Here we use $r_{\rm200}$ and virial radius indifferently and we calculate r$_{\rm200}$ using the scaling relations given by \citet{Arnaud05}  
$r_{200}=1714\frac{(kT/5)^{0.5}}{hz}$ [kpc] where  hz=$\sqrt{((1+z)^3 \Omega_m+\Omega_\Lambda)} $}
on a simulated 300 ks observation of the well studied cluster Abell 1795. 

\noindent Throughout this paper we assume H$_{\rm0}$=70 km s$^{\rm-1}$ and $\Omega_\lambda$=0.73 and $\Omega_{\rm m}$=0.27, which are the default
values in the \texttt{XSPEC}(v12.5) software. 
All errors are quoted at 68\% confidence level for one parameter of interest, unless otherwise specified.   

%%%%%%%%%%%%%%%%%%%%%%%%%%%%%%%%%%%%%%%%%%%%%%%%%%%%%%%%%%%%%%%%%%%%%%%%%%%%%%%%%%%%%%%%%%%%%%%%%%%%%%%%%%%%%%%%%%%%%%%
%
\section{Data reduction and analysis procedures} 
\label{sect:data}

%%%%%%%%%%%%%%%%%%%%%%%%%%%%%%%%%%%
\subsection{The sample}
\begin{figure*}
\begin{tabular}{cc}
\includegraphics[width=8.75cm] {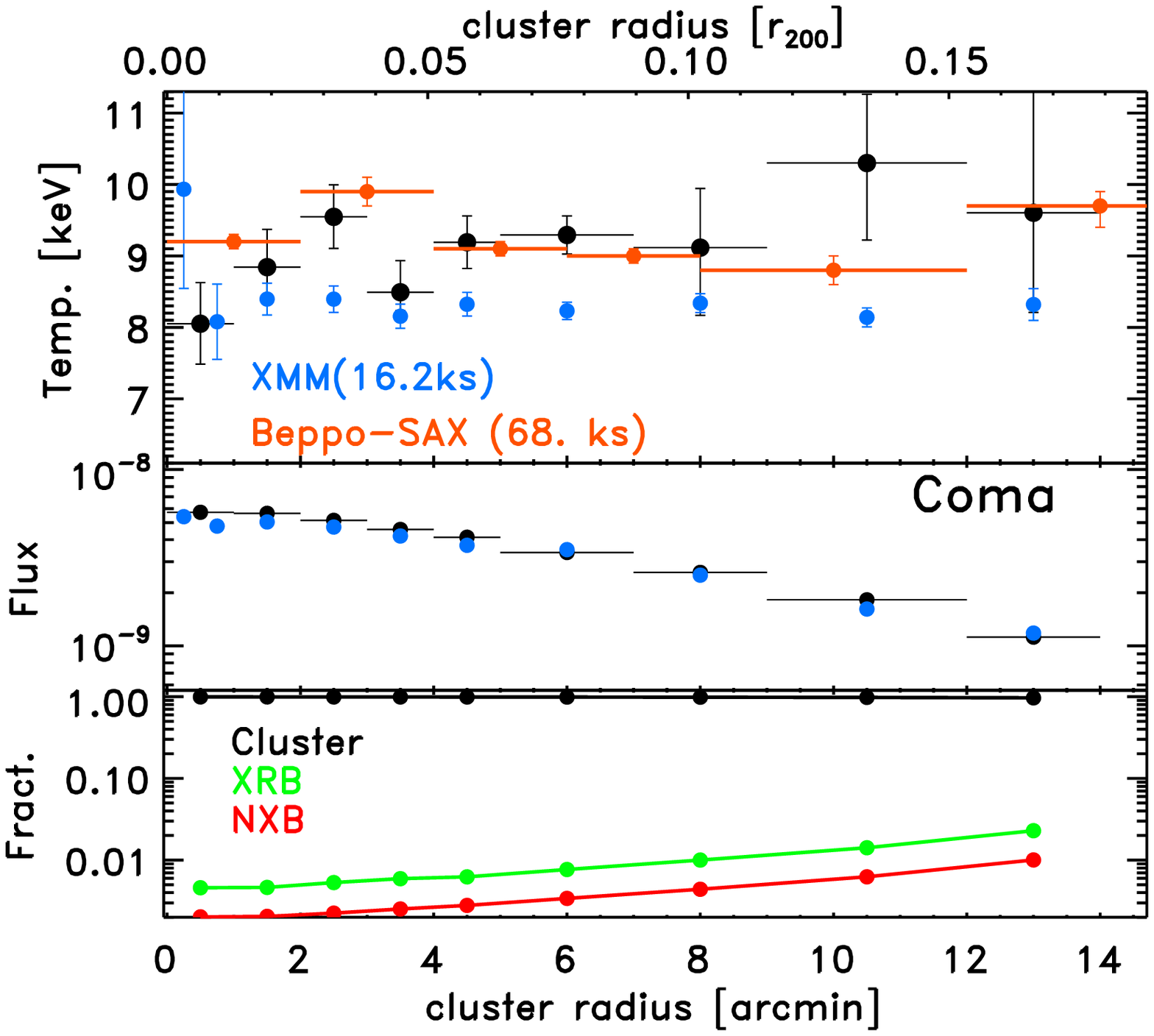}        & \includegraphics[width=8.75cm] {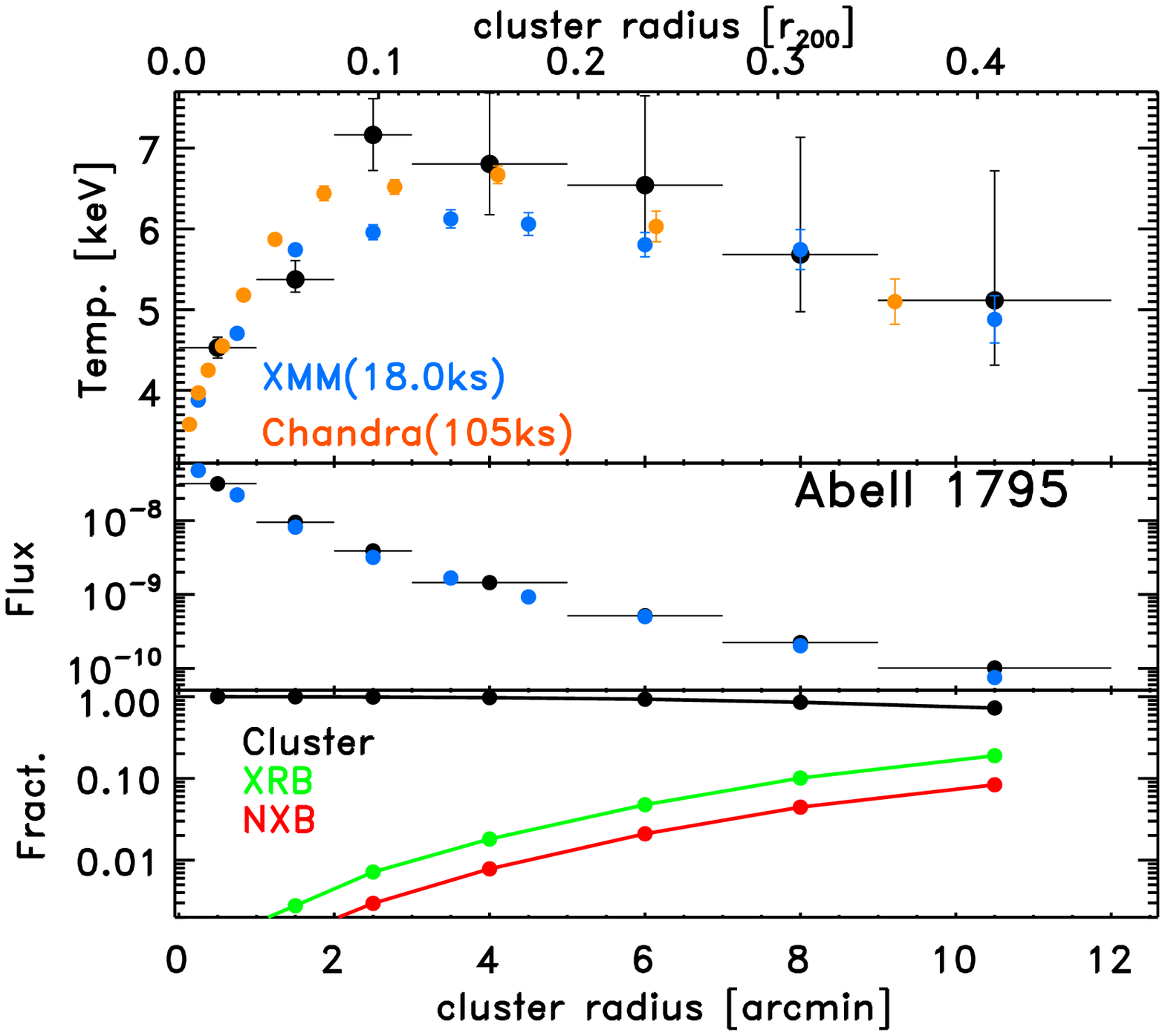} \\
\includegraphics[width=8.75cm] {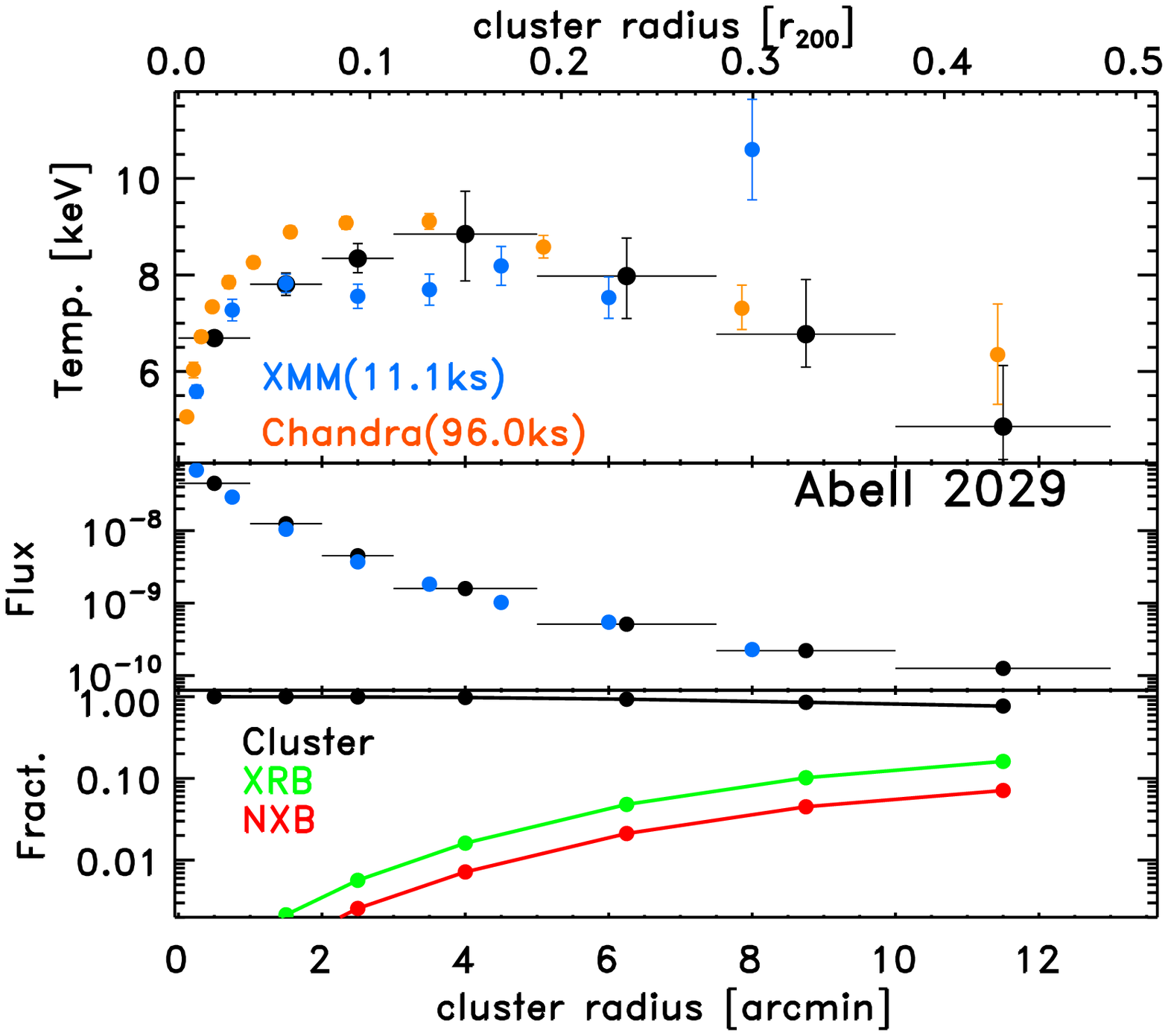}      & \includegraphics[width=8.75cm] {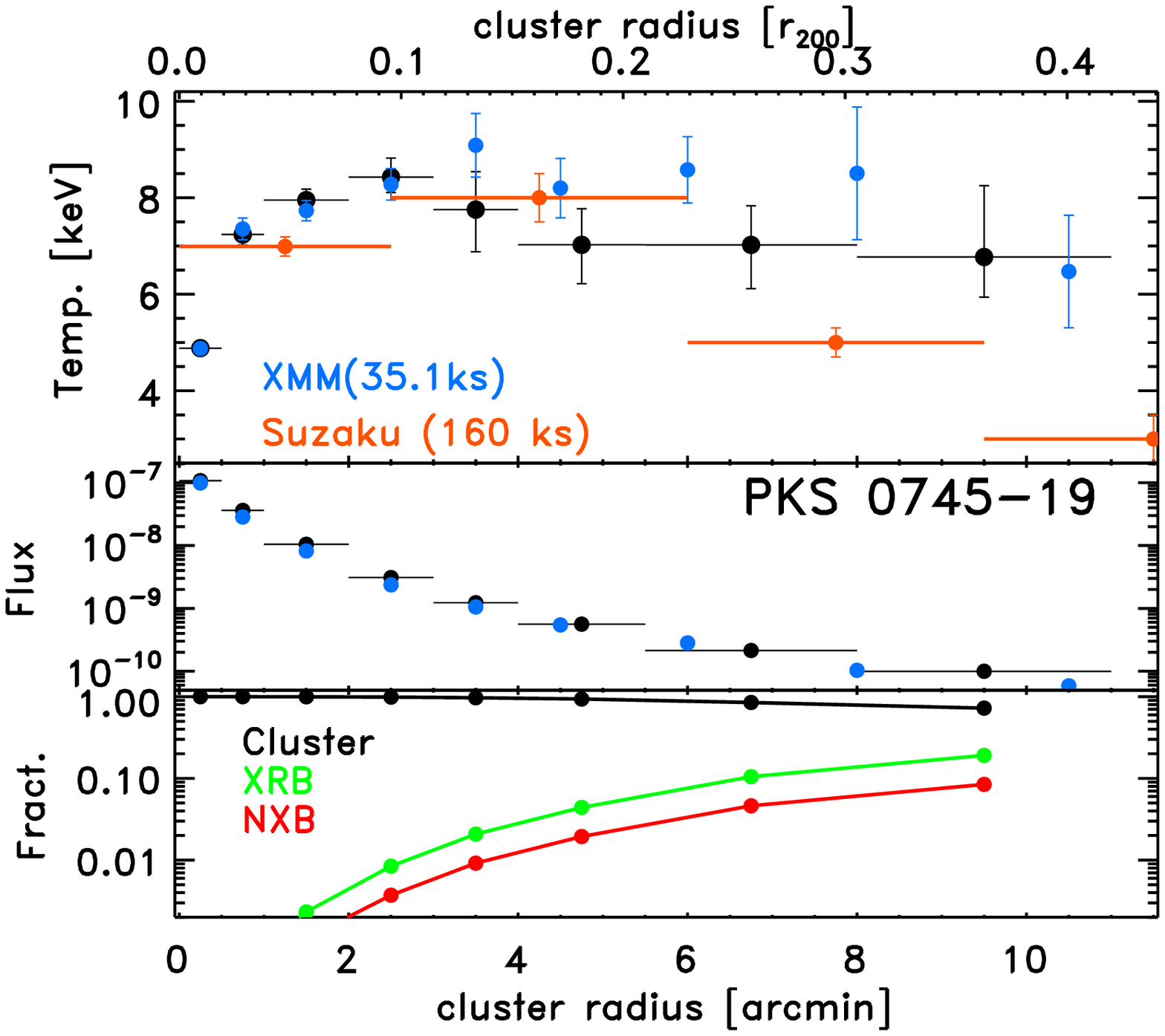}  \\
\includegraphics[width=8.75cm] {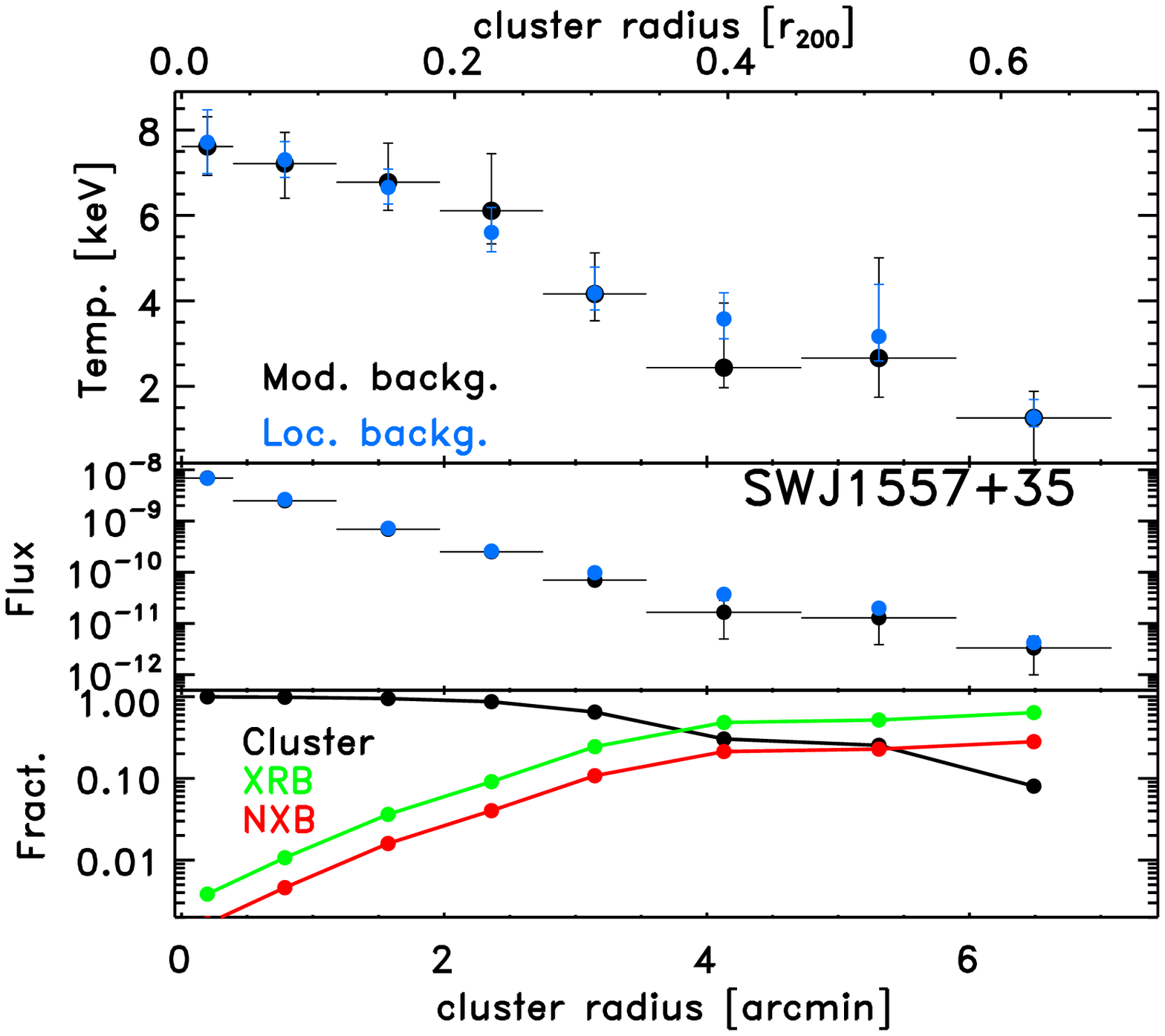}    & \includegraphics[width=8.75cm]{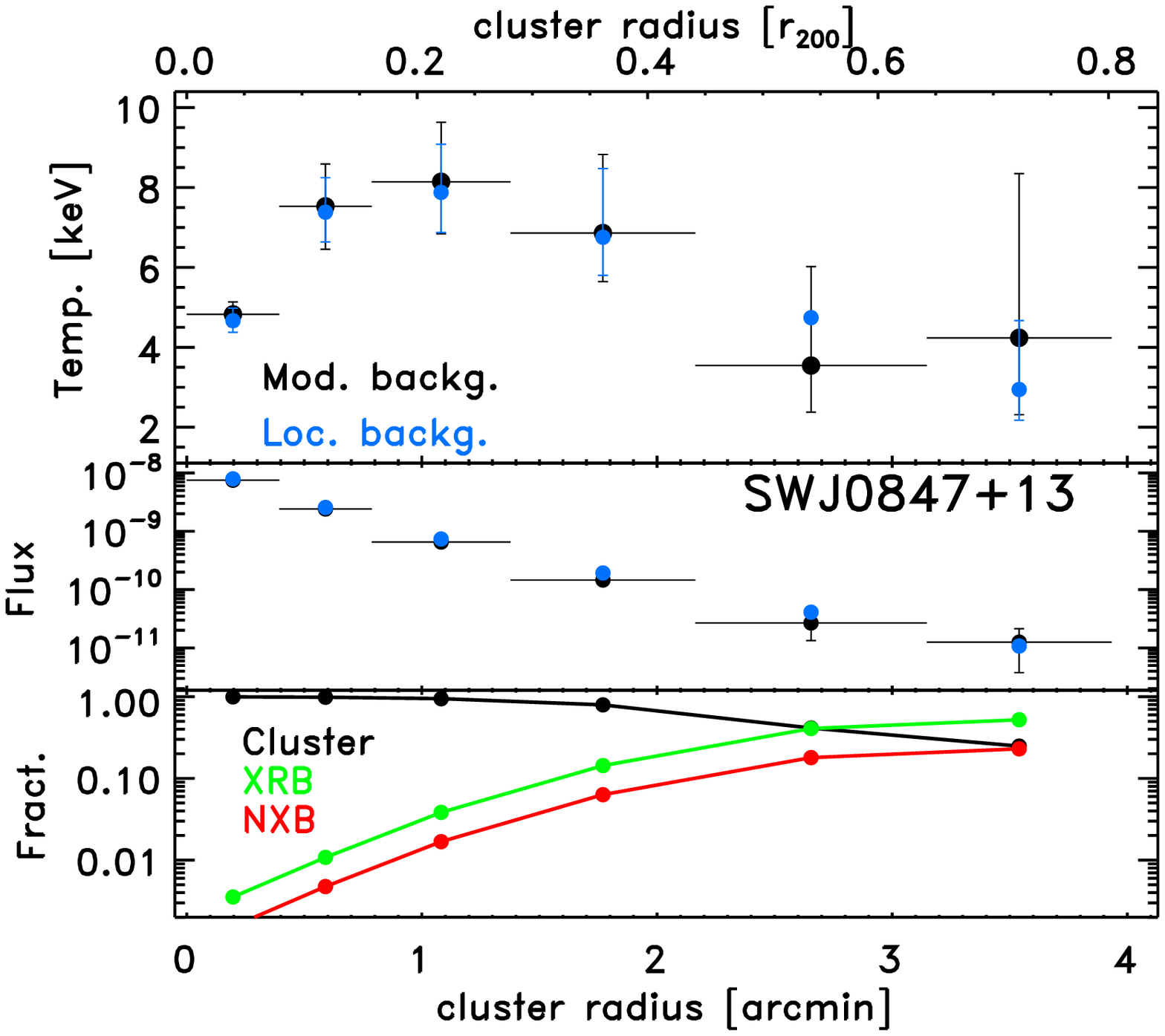}   \\ 
\end{tabular}
\caption{{\bf Upper panels:} The projected temperature profiles of the six clusters of our sample. For the first four  objects, Swift 
XRT measurements (black points) are compared with XMM-Newton (blue points) \cite{Snowden08}. 
In the  Coma plot we also report the BeppoSAX numbers from \cite{Degrandi02}. In the Abell 2029 and Abell 1795 plots 
we report the Chandra measurements from \cite{Vikhlinin06}. In the PKS0745-19 we report the Suzaku points \citep{George09}.
In the case of SWJ1557+35 and SWJ0847+13 we compare measurements obtained with locally evaluated and modeled background.
 {\bf Middle panels:} Fluxes in the 0.3-10. keV  band in cgs deg$^{-2}$ units compared with XMM-Newton values from \cite{Snowden08}, in the first four cases, and with the value
after local background subtraction in the remaining two. 
 {\bf Lower panels:} For each extraction region we plot the relative value of the 0.5-2.0 keV band  flux for the cluster (black), XRB (green) and NXB (red).}
\label{fig:tprof}
\end{figure*}
The X-ray telescope (XRT) on board the Swift satellite \citep{Gehrels04}, uses a Wolter I mirror set, originally designed for
the JET-X telescope \citep{Citterio94}, to focus X-rays (0.2-10 keV) onto a XMM-Newton/EPIC MOS CCD detector \citep{Burrows05}.  
The effective area of the telescope ($\sim$ 150 cm$^2$ at 1.5 keV) is $\sim$ 3. smaller than one single XMM-Newton MOS.
The PSF, similar to XMM-Newton, is characterized by an half energy width (HEW) of $\sim$ 18\arcsec  at 1.5 keV \citep{Moretti05}.
As part of the Swift scientific payload, since the start of the mission (November 2004), Swift-XRT has been mostly used to observe 
GRB afterglows and other variable sources. Galaxy clusters are observed mostly for calibration purposes or serendipitously.

\noindent We cross-correlated the XRT archive updated on April 2010 with the BAX database (\texttt{http://bax.ast.obs-mip.fr/}) 
searching for cluster observed with more than 20,000 events registered in the 0.7-7.0 keV energy band. The sample consists of four 
famous and very well studied objects (Fig~\ref{fig:imaprof},Tab.~\ref{tab:obse}). For these clusters the XRT archive data allowed us to 
measure the projected temperature profile as far as the edge of the telescope field of view corresponding to 0.2-0.6  r$_{200}$.

\noindent We added 2 more clusters to this sample; these are SWJ1557+3530  and SWJ0847+1331 at redshift z=0.15 and z=0.36 respectively
(according to X-ray spectral analysis, see below ), which were serendipitously observed in the field of the follow-up observations of  GRB090409 and 
GRB051016B and are the two  highest signal objects of the SXCS catalog \citep[Moretti et al. in preparation]{Moretti07}.  
They are already cataloged as MaxBCGJ239.42665+35.50827 \citep{Koester07} (photoz=0.155) and 
as WHL J084749.3+133140 \citep{Wen09} (z=0.36) respectively. 
They are X-ray bright (L$_{X}\sim$ 3 and 7 10$^{44}$ erg s$^-1$  in the 0.5-2.0 keV rest frame energy band) with average temperatures 
of 6.0 and 6.8 keV respectively.  They can be considered rich also form the point of view of the optical classification having  
35 and 42  galaxies associated according the MaxBCG and the WHL catalogs respecitvely. The cataloged redshift measurements are in 
very good agreement with ours (see Tab.~\ref{tab:obse}).
 Differently from the first four, these two objects do not entirely fill the field of view  
 allowing us to perform a consistency check of  our background model through a comparison with  the local background (see below).
%%%%%%%%%%%%%%%%%%%%%%%%%%%%%%%%%%%
\subsection{Data Reduction}\label{sect:dred}
\noindent
Data reduction was performed using the standard software (HEADAS software, v6.8, CALDB version 20091130, Nov 2009) and 
following the procedures described in the instrument user guide \footnote{http://heasarc.nasa.gov/docs/swift/analysis/documentation}.  
At variance with these, we excluded the external (Detx $>$ 90 and Detx $<$510 ) CCD columns which are affected by the presence of 
out-of-time-events from corner calibration sources \citep[see][for a detailed map of the CCD and a discussion on the XRT background]{Moretti09}. 
This left us with a nominal field of view of 16.5$\arcmin~\times~$18.9$\arcmin$ (0.087 deg$^2$).
Different observations of the same objects and relative exposure maps were merged  by means of the the \texttt{extractor} and \texttt{farith} tasks 
of the HEADAS software respectively.

\noindent
Before performing the spectral analysis, we ran the \texttt{CIAO wavedetect} and eliminated  all the events within the circles centered on the  detected
source positions with radius such that the PSF surface brightness equals the background (typically $\lesssim$ 10 pixels $\lesssim$ 23\arcsec). 
Vignetted and non-vignetted exposure maps were built consequently accounting for the excised regions.

%%%%%%%%%%%%%%%%%%%%%%%%%%%%%%%%%%%
\subsection{Spectral Analysis}
\label{sect:spectra}
\begin{figure}
\includegraphics[width=\columnwidth] {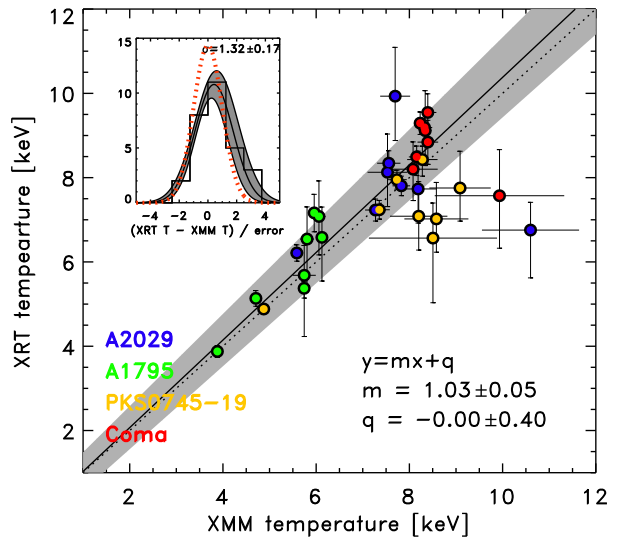}     
\caption{Swift-XRT and XMM-Newton temperature measurement comparison. The continuous line is the linear fit to the data, while the dotted 
line is the expected distribution. In the inset the scatter of the measurement is shown with the thick black line. The best fit with its uncertainties 
is shown with thin black lines and grey area, while the expected scatter is drawn with red dotted line. }
\label{fig:specx}
\end{figure}
\noindent
We measured the temperature profiles, binning the  data in projected annuli of  amplitude1.0$\arcmin$-2.5$\arcmin$ moving from central to 
the external part of the image (Fig~\ref{fig:imaprof}). The statistics of the extracted spectra strongly vary going from the center, where the 
extraction region size is determined only by the PSF toward the outskirts of the sources, where the size of the regions is determined  by ensuring 
a minimum of 200  source counts in the 0.7-2.0 keV energy band.   
Because the XRT standard software does not include any specific task for the spectral analysis of the extended sources, we developed a procedure made by combination of ad-hoc \texttt{IDL} routines and \texttt{Ftools} scripts to calculate the vignetting and the exposure corrections and to estimate the background.

\noindent For the vignetting calculation, we, first, created a vignetting map at 1.5 keV, dividing the  exposure map by the un-vignetted exposure. 
Then, we  created the maps at different energies, using the vignetting analytical description given by \cite{Moretti04}. 
Finally  we used the exposure and vignetting maps weighted by the cluster surface brightness profile to calculate the 
effective time and the vignetting correction, for each annulus. As the vignetting depends on energy, in order to take 
into account for these corrections in the spectral analysis we opportunely modified the nominal  ARF file.  

\noindent The evaluation of the background is surely the most important and delicate step in the procedure. This is primarily because 
the virial regions (R$_{\rm200} >$ 15$\arcmin$) of nearby massive clusters subtend the entire field of view. Therefore, the
background cannot be estimated locally.
For our purposes we consider the total XRT background as the sum of an instrument (NXB) and a cosmic (XRB) component. 
The latter one is, in turn, the sum of a galactic (GXRB) and extragalactic (CXRB) component, with very different  spectral and spatial characteristics.

\noindent We evaluated the instrumental background (NXB) using the data collected, during the observations, in the regions of the detector which are 
not exposed to the sky (NESR, Not-Exposed-Sky-Regions). These are four different small regions (2507 pixels each) close to the CCD boundary and 
delimited by the field of view and the corner sources \citep{Moretti09}. 

\noindent 
To estimate the XRB we used a statistical approach: we made use of a large number (135) of deep GRB follow-up observations 
to study the XRB characteristics. Because GRB are uniformly distributed,  these are deep exposures on random positions of the sky,
totally uncorrelated with already known bright X-ray sources, and provide us with a data set which is very well suited to statistically 
characterize the XRB. In the following we will refer to this data set as the blank fields (BFs). 
A similar data-set was used by \cite{Moretti09}  to perform an absolute measurement of the CXRB  in the 1.5-7 keV range.
Here, we found  that all the BF spectra, in the 0.7-7.0 keV energy band can be well described by the sum of a bremsstrahlung and 
an absorbed power law, representing the GXRB and CXRB respectively,  

\noindent We defer the detailed description of the instrumental  and cosmic background analysis 
together with  the full evaluation of the systematic errors in temperature measurements to the next two sections 
(Sect.~\ref{sect:backg},~\ref{sect:ssist}), while here we focus on the remaining details of the analysis and the results.

\noindent As high energy limit we use 7.0 keV, because, between 7 and 10 keV, the instrumental background is significantly higher 
due to the presence of Nickel and Gold lines. We set the  low energy  limit at 0.7 
keV both because around 0.5 keV response matrix calibrations are more problematic due to the presence of uncorrected charge traps 
\citep{Godet09}, and because we found that, below 0.7 keV, the XRB is less reproducible due to the presence of some extra local 
components (see also \cite{Kuntz00}). In the case of Coma, both brightness and temperature are so high that we easily extended our 
analysis up to 10.0 keV. In the case of  PKS0745-19, which is on the Galactic plane, we used the 1.0-7.0 keV band  to reduce the 
uncertainty on the background estimate which is higher at lower Galactic latitudes. 

\noindent 
We used \texttt{XSPEC}(v12.5) Cash statistics to fit the cluster+XRB spectra from the annular regions
with a \texttt{bremss}+\texttt{wabs}$\times$(\texttt{pow}+\texttt{apec}) model, where the 
absorbed APEC model represents the cluster emission, using the NESR spectra as background. 
We used literature red-shift values (\texttt{http://bax.ast.obs-mip.fr/}) for  all the clusters but for 
SWJ1557+3530 and SWJ0847+1331 which are not yet cataloged. In these cases we measured the redshift 
using the Fe-K line in the X-ray spectrum from the central 1$\arcmin$ radius.
We left the metallicity parameter free to vary only when the estimated counts of the sources exceed 5,000, 
freezing it to the 0.3 Z$_{\odot}$ in the other cases. 
 Concerning the Galactic absorption, for each cluster, we took advantage of the good statistics in the 
three central annuli, and  we let the N$_H$ absorbing column vary in the fit. 
As we found, in all the cases, a $<$2$\sigma$ consistency of the best fit value with the value derived from the 
HI Galaxy map  \citep{Kalberla05}, we froze the Galactic N$_H$ absorbing columns to these values. 
In three cases these estimate coincide with the XMM best fit values (Coma, Abell 1795, PKS0745-19 differences $<$3\%)
\citep{Snowden08}, while, in the case of Abell 2029, the difference is $\sim$20\%.

\noindent
We tried to fit the cluster data together with XRB, but 
the statistics of the data did not allow us to constrain the XRB parameters together with the cluster ones,
even when the XRB is a significant fraction of the total signal. 
Instead, for all our spectra, we froze the four XRB  parameters to the median values of the BF distributions, 
just re-normalizing them by the area of the extraction region.   
We calculated the systematic uncertainties of this approach reproducing the same procedure on a large 
sample of simulated spectra (Sect. \ref{sect:ssist}).
Finally we calculated the total error as the quadratic sum of the statistical error (the one provided by \texttt{XSPEC}) 
and the systematic one. 
The results are shown in the Fig.~\ref{fig:tprof} and reported in details in Tab.~\ref{tab:dataprof}.  

\noindent
We note that in the case of the PKS0745-19 the systematic errors are probably underestimated, due to the fact 
that this cluster lies on the Galactic plane,  whereas our XRB statistical analysis is suitable for the extragalactic sky 
(Galactic latitude $>$ 20$^{\circ}$).
We mitigated this problem limiting the spectral analysis for this particular cluster to energies higher than 1.0 keV.   

\noindent
In Fig. \ref{fig:specx} we plot the XRT temperature measurements of the first four clusters, compared with XMM-Newton results
from \cite{Snowden08}: we generally found  a good agreement between the two instruments without any appreciable bias or 
systematic trend as shown by the linear fit.  On the other hand, calculating the scatter as the variance of the (XRT)-(XMM-Newton) 
difference distribution, divided by the errors, we measured a scatter which is slightly larger than the expected: 1.32$\pm$0.17 
instead of 1.0. 
Most of this scatter can be ascribed to the outer regions of  the cluster PKS0745-19, for which, as said, the errors 
are underestimated. Indeed excluding them lowers the scatter to 1.20$\pm$0.18, consistent with the expectation at 1.1 $\sigma$
level. We obtained the same result (1.22$\pm$0.20) artificially increasing the PKS0745-19 errors by a factor 1.4.
In this comparison we performed  the spectral analysis exactly in the same regions used by \citep{Snowden08}, 
while data plotted in Fig.~\ref{fig:tprof} and listed 
in Tab.~\ref{tab:dataprof} are in (slightly) different regions, modified in order to optimize XRT data statistics. 

\begin{figure}
\includegraphics[width=\columnwidth]   {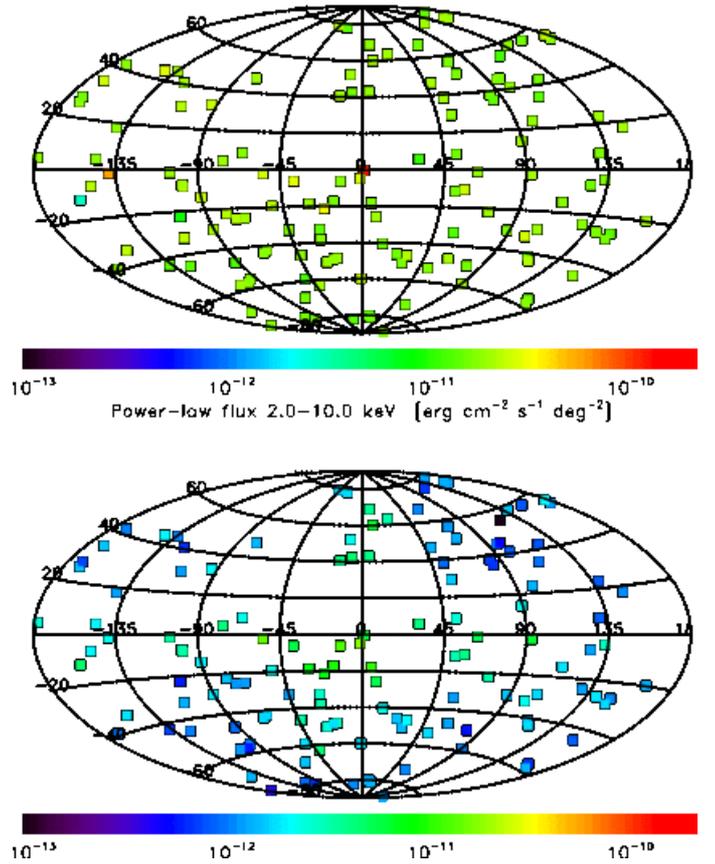}      \\ 
\caption{ The Aitoff projection of the Galactic coordinates of the non resolved emission in the blank fields (BFs). 
{\bf Upper Panel:} the CXRB flux in the 2.0-10.0 keV band. {\bf Lower Panel:} the GXRB flux in the 0.5-2.0 keV band.}  
\label{fig:bckmap}
\end{figure}
\noindent
Due to small apparent size and the good statistics SWJ1557+3530 and SWJ0847+1331 are the best cases, 
in the XRT archive, to perform a consistency check between temperature measurement performed with a locally 
evaluated and a background model: results are in good agreement, as shown in the last two panels of Fig.~\ref{fig:tprof}.  
\begin{table*}
\begin{center}
\caption[]{Extraction regions and spectral analysis results. For each annulus we report the radii, the area, the effective exposure time accounting for 
exposure maps and vignetting, the mean cluster surface brightness, the total events, the source/total ratio and the best fits values. 
The statistical error contribution to the total temperature errors are reported in parenthesis. }
\begin{tabular}{|l|cccccc|cc|}
\hline
Cluster  & ann. rad.      & area        & exp.  & SB$_{[0.5-2.]}$                                       & tot. evt.$_{[0.7-2.]}$& S/(S+B)  &      kT     & Z  \\
            & [$\arcmin$]  & [deg$^2$]& [ks]   & [\small{erg s$^{-1}$ cm$^{-2}$ deg$^{-2}$}] &              & [\%]       &      [keV] & Z$_{\odot}$ \\
\hline
\hline
      Coma &  0.0  - 1.0 &  8.7E-04 &  41.9 &  5.72E-09$\pm$  3.04E-10 &    3.1e+03 &  99.5 &  8.1$_{- 0.6( 0.6)}^{+ 0.6  ( 0.6)}$ & 0.30$_{ 0.00}^{+0.00}$ \\
      Coma &  1.0  - 2.0 &  2.5E-03 &  41.6 &  5.64E-09$\pm$  2.89E-10 &    8.6e+03 &  99.4 &  8.8$_{- 0.5( 0.5)}^{+ 0.5  ( 0.5)}$ & 0.24$_{-0.08}^{+0.08}$ \\
      Coma &  2.0  - 3.0 &  4.0E-03 &  41.8 &  5.14E-09$\pm$  2.61E-10 &    1.2e+04 &  99.4 &  9.5$_{- 0.4( 0.4)}^{+ 0.4  ( 0.4)}$ & 0.22$_{-0.07}^{+0.07}$ \\
      Coma &  3.0  - 4.0 &  6.1E-03 &  42.0 &  4.57E-09$\pm$  2.31E-10 &    1.7e+04 &  99.3 &  8.5$_{- 0.3( 0.3)}^{+ 0.4  ( 0.4)}$ & 0.20$_{-0.05}^{+0.05}$ \\
      Coma &  4.0  - 5.0 &  7.5E-03 &  41.8 &  4.13E-09$\pm$  2.09E-10 &    1.9e+04 &  99.2 &  9.2$_{- 0.4( 0.4)}^{+ 0.4  ( 0.4)}$ & 0.25$_{-0.05}^{+0.06}$ \\
      Coma &  5.0  - 7.0 &  1.9E-02 &  40.5 &  3.38E-09$\pm$  1.70E-10 &    3.8e+04 &  99.0 &  9.3$_{- 0.3( 0.3)}^{+ 0.3  ( 0.3)}$ & 0.28$_{-0.04}^{+0.04}$ \\
      Coma &  7.0  - 9.0 &  2.6E-02 &  32.3 &  2.61E-09$\pm$  1.31E-10 &    3.3e+04 &  98.8 &  9.1$_{- 0.7( 0.3)}^{+ 0.8  ( 0.3)}$ & 0.19$_{-0.04}^{+0.05}$ \\
      Coma &  9.0  -12.0 &  5.3E-02 &  16.5 &  1.82E-09$\pm$  9.20E-11 &    2.2e+04 &  98.2 & 10.3$_{- 0.8( 0.4)}^{+ 1.0  ( 0.4)}$ & 0.45$_{-0.08}^{+0.08}$ \\
      Coma & 12.0  -14.0 &  2.9E-02 &   5.6 &  1.12E-09$\pm$  6.03E-11 &    2.7e+03 &  97.1 &  9.6$_{- 1.2( 1.0)}^{+ 1.7  ( 1.5)}$ & 0.30$_{ 0.00}^{+0.00}$ \\
\hline
 Abell2029 &  0.0  - 1.0 &  8.7E-04 &  38.7 &  4.49E-08$\pm$  2.26E-09 &    2.4e+04 &  99.9 &  6.7$_{- 0.1( 0.1)}^{+ 0.1  ( 0.1)}$ & 0.58$_{-0.05}^{+0.05}$ \\
 Abell2029 &  1.0  - 2.0 &  2.6E-03 &  39.4 &  1.25E-08$\pm$  6.29E-10 &    2.0e+04 &  99.8 &  7.8$_{- 0.2( 0.2)}^{+ 0.2  ( 0.2)}$ & 0.36$_{-0.05}^{+0.05}$ \\
 Abell2029 &  2.0  - 3.0 &  4.4E-03 &  40.6 &  4.50E-09$\pm$  2.29E-10 &    1.2e+04 &  99.3 &  8.3$_{- 0.3( 0.3)}^{+ 0.3  ( 0.3)}$ & 0.43$_{-0.06}^{+0.07}$ \\
 Abell2029 &  3.0  - 5.0 &  1.2E-02 &  41.2 &  1.59E-09$\pm$  8.07E-11 &    1.2e+04 &  98.1 &  8.8$_{- 0.7( 0.4)}^{+ 0.9  ( 0.5)}$ & 0.27$_{-0.06}^{+0.07}$ \\
 Abell2029 &  5.0  - 7.5 &  2.7E-02 &  40.0 &  5.12E-10$\pm$  2.62E-11 &    8.9e+03 &  94.4 &  8.0$_{- 0.7( 0.4)}^{+ 0.8  ( 0.4)}$ & 0.22$_{-0.07}^{+0.08}$ \\
 Abell2029 &  7.5  -10.0 &  3.8E-02 &  28.9 &  2.21E-10$\pm$  1.17E-11 &    4.4e+03 &  88.5 &  6.8$_{- 0.6( 0.4)}^{+ 1.0  ( 0.8)}$ & 0.30$_{ 0.00}^{+0.00}$ \\
 Abell2029 & 10.0  -13.0 &  4.3E-02 &   9.6 &  1.26E-10$\pm$  7.66E-12 &    1.2e+03 &  83.5 &  4.9$_{- 0.7( 0.6)}^{+ 0.9  ( 0.8)}$ & 0.30$_{ 0.00}^{+0.00}$ \\
\hline
 Abell1795 &  0.0  - 1.0 &  8.7E-04 &  20.0 &  3.16E-08$\pm$  1.61E-09 &    1.0e+04 &  99.9 &  4.5$_{- 0.1( 0.1)}^{+ 0.1  ( 0.1)}$ & 0.46$_{-0.06}^{+0.07}$ \\
 Abell1795 &  1.0  - 2.0 &  2.6E-03 &  19.8 &  9.50E-09$\pm$  4.86E-10 &    8.7e+03 &  99.7 &  5.4$_{- 0.2( 0.2)}^{+ 0.2  ( 0.2)}$ & 0.36$_{-0.06}^{+0.07}$ \\
 Abell1795 &  2.0  - 3.0 &  4.4E-03 &  19.2 &  3.90E-09$\pm$  2.02E-10 &    5.2e+03 &  99.3 &  7.2$_{- 0.4( 0.4)}^{+ 0.4  ( 0.4)}$ & 0.30$_{ 0.00}^{+0.00}$ \\
 Abell1795 &  3.0  - 5.0 &  1.4E-02 &  19.2 &  1.45E-09$\pm$  7.48E-11 &    6.4e+03 &  98.0 &  6.8$_{- 0.5( 0.3)}^{+ 0.7  ( 0.4)}$ & 0.30$_{ 0.00}^{+0.00}$ \\
 Abell1795 &  5.0  - 7.0 &  1.9E-02 &  17.8 &  5.16E-10$\pm$  2.76E-11 &    3.1e+03 &  94.8 &  6.5$_{- 0.6( 0.5)}^{+ 0.8  ( 0.6)}$ & 0.30$_{ 0.00}^{+0.00}$ \\
 Abell1795 &  7.0  - 9.0 &  2.7E-02 &  15.9 &  2.23E-10$\pm$  1.26E-11 &    1.9e+03 &  89.3 &  5.7$_{- 0.6( 0.6)}^{+ 0.7  ( 0.7)}$ & 0.30$_{ 0.00}^{+0.00}$ \\
 Abell1795 &  9.0  -12.0 &  4.0E-02 &  11.4 &  1.01E-10$\pm$  6.41E-12 &    1.0e+03 &  79.9 &  5.1$_{- 0.8( 0.7)}^{+ 1.2  ( 1.1)}$ & 0.30$_{ 0.00}^{+0.00}$ \\
\hline
PKS0745-19 &  0.0  - 0.5 &  2.2E-04 &  75.6 &  1.08E-07$\pm$  5.42E-09 &    3.3e+04 & 100.0 &  4.9$_{- 0.1( 0.1)}^{+ 0.1  ( 0.1)}$ & 0.44$_{-0.04}^{+0.04}$ \\
PKS0745-19 &  0.5  - 1.0 &  6.5E-04 &  77.2 &  3.62E-08$\pm$  1.82E-09 &    2.9e+04 &  99.9 &  7.2$_{- 0.2( 0.2)}^{+ 0.2  ( 0.2)}$ & 0.37$_{-0.04}^{+0.04}$ \\
PKS0745-19 &  1.0  - 2.0 &  2.4E-03 &  77.1 &  1.05E-08$\pm$  5.28E-10 &    3.0e+04 &  99.7 &  8.0$_{- 0.2( 0.2)}^{+ 0.2  ( 0.2)}$ & 0.38$_{-0.04}^{+0.04}$ \\
PKS0745-19 &  2.0  - 3.0 &  4.1E-03 &  77.8 &  3.09E-09$\pm$  1.57E-10 &    1.5e+04 &  99.0 &  8.4$_{- 0.3( 0.3)}^{+ 0.4  ( 0.4)}$ & 0.43$_{-0.06}^{+0.06}$ \\
PKS0745-19 &  3.0  - 4.0 &  6.1E-03 &  77.9 &  1.23E-09$\pm$  6.28E-11 &    9.4e+03 &  97.6 &  7.8$_{- 0.7( 0.4)}^{+ 0.8  ( 0.4)}$ & 0.30$_{ 0.00}^{+0.00}$ \\
PKS0745-19 &  4.0  - 5.5 &  1.2E-02 &  74.9 &  5.61E-10$\pm$  2.88E-11 &    8.5e+03 &  95.1 &  7.0$_{- 0.6( 0.4)}^{+ 0.8  ( 0.5)}$ & 0.30$_{ 0.00}^{+0.00}$ \\
PKS0745-19 &  5.5  - 8.0 &  2.9E-02 &  68.8 &  2.14E-10$\pm$  1.10E-11 &    7.8e+03 &  88.1 &  7.0$_{- 0.8( 0.6)}^{+ 0.8  ( 0.6)}$ & 0.30$_{ 0.00}^{+0.00}$ \\
PKS0745-19 &  8.0  -11.0 &  4.9E-02 &  40.9 &  9.97E-11$\pm$  5.37E-12 &    4.2e+03 &  77.8 &  6.8$_{- 0.9( 0.6)}^{+ 1.5  ( 1.2)}$ & 0.30$_{ 0.00}^{+0.00}$ \\
\hline
SWJ1557+35 &  0.0  - 0.4 &  1.3E-04 & 187.3 &  6.91E-09$\pm$  3.70E-10 &    2.7e+03 &  99.6 &  7.6$_{- 0.7( 0.7)}^{+ 0.7  ( 0.7)}$ & 0.30$_{ 0.00}^{+0.00}$ \\
SWJ1557+35 &  0.4  - 1.2 &  1.1E-03 & 176.3 &  2.49E-09$\pm$  1.28E-10 &    7.8e+03 &  98.9 &  7.2$_{- 0.6( 0.4)}^{+ 0.7  ( 0.4)}$ & 0.30$_{ 0.00}^{+0.00}$ \\
SWJ1557+35 &  1.2  - 2.0 &  2.2E-03 & 170.1 &  6.90E-10$\pm$  3.62E-11 &    4.4e+03 &  96.1 &  6.8$_{- 0.6( 0.4)}^{+ 0.7  ( 0.5)}$ & 0.30$_{ 0.00}^{+0.00}$ \\
SWJ1557+35 &  2.0  - 2.8 &  3.1E-03 & 163.3 &  2.51E-10$\pm$  1.37E-11 &    2.5e+03 &  90.4 &  6.1$_{- 0.7( 0.6)}^{+ 0.8  ( 0.6)}$ & 0.30$_{ 0.00}^{+0.00}$ \\
SWJ1557+35 &  2.8  - 3.5 &  4.3E-03 & 154.5 &  7.00E-11$\pm$  4.35E-12 &    1.3e+03 &  75.9 &  4.2$_{- 0.6( 0.4)}^{+ 0.9  ( 0.8)}$ & 0.30$_{ 0.00}^{+0.00}$ \\
SWJ1557+35 &  3.5  - 4.7 &  7.6E-03 & 123.5 &  1.66E-11$\pm$  1.40E-12 &    8.7e+02 &  49.7 &  2.4$_{- 0.4( 0.3)}^{+ 0.7  ( 0.5)}$ & 0.30$_{ 0.00}^{+0.00}$ \\
SWJ1557+35 &  4.7  - 5.9 &  1.0E-02 & 101.0 &  1.29E-11$\pm$  1.23E-12 &    8.5e+02 &  42.4 &  2.7$_{- 0.6( 0.5)}^{+ 1.1  ( 0.6)}$ & 0.30$_{ 0.00}^{+0.00}$ \\
SWJ1557+35 &  5.9  - 7.1 &  1.3E-02 &  87.6 &  3.32E-12$\pm$  6.12E-13 &    6.9e+02 &  21.6 &  1.3$_{- 0.3( 0.2)}^{+ 1.2  ( 0.5)}$ & 0.30$_{ 0.00}^{+0.00}$ \\
\hline
SWJ0847+13 &  0.0  - 0.4 &  1.3E-04 & 104.9 &  7.50E-09$\pm$  4.08E-10 &    2.2e+03 &  99.7 &  4.8$_{- 0.3( 0.3)}^{+ 0.3  ( 0.3)}$ & 0.30$_{ 0.00}^{+0.00}$ \\
SWJ0847+13 &  0.4  - 0.8 &  4.0E-04 & 100.9 &  2.41E-09$\pm$  1.34E-10 &    1.7e+03 &  98.9 &  7.5$_{- 0.9( 0.8)}^{+ 1.1  ( 0.8)}$ & 0.30$_{ 0.00}^{+0.00}$ \\
SWJ0847+13 &  0.8  - 1.4 &  1.1E-03 &  94.7 &  6.53E-10$\pm$  3.82E-11 &    1.2e+03 &  95.9 &  8.1$_{- 1.2( 1.0)}^{+ 1.5  ( 1.3)}$ & 0.30$_{ 0.00}^{+0.00}$ \\
SWJ0847+13 &  1.4  - 2.2 &  2.4E-03 &  84.4 &  1.46E-10$\pm$  1.00E-11 &    6.2e+02 &  84.8 &  6.9$_{- 1.2( 1.1)}^{+ 2.0  ( 1.8)}$ & 0.30$_{ 0.00}^{+0.00}$ \\
SWJ0847+13 &  2.2  - 3.1 &  4.6E-03 &  73.2 &  2.67E-11$\pm$  2.71E-12 &    3.7e+02 &  58.5 &  3.5$_{- 1.0( 0.9)}^{+ 1.8  ( 1.6)}$ & 0.30$_{ 0.00}^{+0.00}$ \\
SWJ0847+13 &  3.1  - 3.9 &  4.0E-03 &  85.9 &  1.26E-11$\pm$  2.19E-12 &    2.5e+02 &  37.6 &  4.2$_{- 1.8( 1.5)}^{+ 3.7  ( 3.2)}$ & 0.30$_{ 0.00}^{+0.00}$ \\
\hline
\end{tabular}
\label{tab:dataprof} 
\end{center}
\end{table*}
%
%%%%%%%%%%%%%%%%%%%%%%%%%%%%%%%%%%%%%%%%%%%%%%%%%%%%%%%%%%%%%%%%%%%%%%%%%%%%%%%%%%%%%%%%%%%%%%%%%%%%%%%%%%%%%%%%%%%%%%%
\section{Background Analysis}
\label{sect:backg} 
To study both the NXB and XRB we made use of a large sample of blank fields (BFs).
We selected the 135 (31 at low Galactic latitude) GRB follow up observations from January 2006 to April 2009 with a nominal exposure time longer 
than 10 ks and shorter than 300 ks (Fig.~\ref{fig:bckmap}).
We reduced these data and eliminated detected sources, following the same procedure we used for cluster data and described in Sect. \ref{sect:spectra}.
\subsection{Instrumental background}
\label{sect:nxb}
\begin{figure}
\includegraphics[width=\columnwidth]   {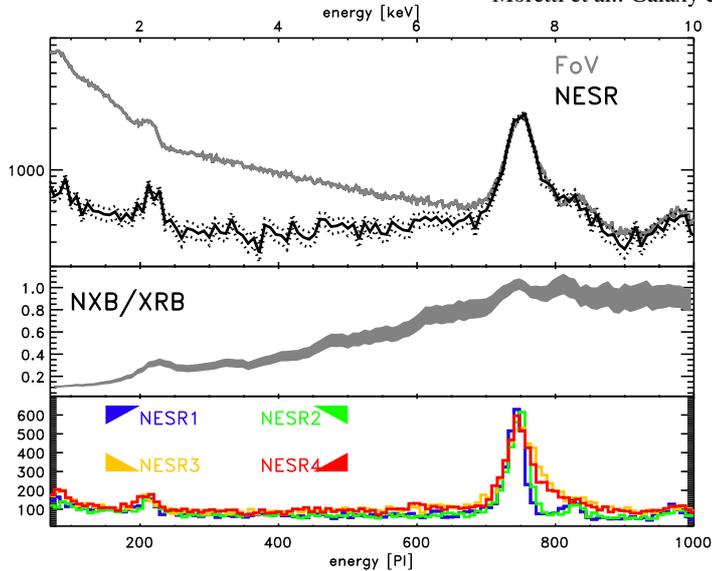}      \\ 
\caption{{\bf Upper panel:} The raw spectrum (P.I. channel distribution; 1 PI$\sim$ 10 eV ) of a large number of blank fields (BF, in grey) 
compared with the spectrum registered in the NESR regions in black. {\bf Middle panel:} The ratio NESR/BF gives the relative importance 
of the NXB as function of energy in absence of bright sources. {\bf Bottom panel:}  The 4 NESR spectra show not negligible differences. }
\label{fig:nxb_spec}
\end{figure}
\begin{figure}
\includegraphics[width=\columnwidth]   {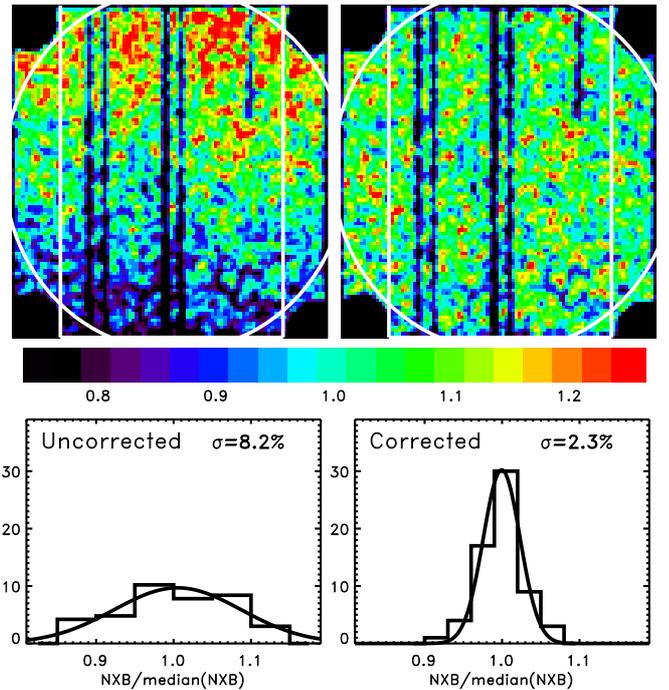}      \\ 
\caption{{\bf Upper-left:} The CCD map in the 7.0-10.0 keV energy band from 6.3 Ms of BFs, normalized to the central raw to show the NXB vertical gradient; 
{\bf Lower-left:} the residual distribution calculated in 50$\times$50 pixel cells. 
{\bf Upper-right:}  The CCD map in the 7.0-10.0 keV energy band from 6.3 Ms of BFs, normalized to the central raw to show the NXB vertical gradient,
corrected by the linear approximation.
{\bf Lower-left:} the residual distribution calculated in 50$\times$50 pixel cells with the linear gradient correction applied. }
\label{fig:nxbgrad}
\end{figure}

In the Swift-XRT CCD, we have the opportunity of estimating the NXB directly from the signal registered in the NESR during the observations.
This method allows to optimally map the NXB  time variations which occur due 
to changes in the satellite environment at any time scales from minutes to years and that are very difficult to be accounted for  otherwise.   
Indeed, as shown in the Fig.~\ref{fig:nxb_spec}, the raw spectrum registered in the NESR regions, once re-normalized for the area, accurately
reproduces the signal registered in whole the field of view beyond 7 keV, where the instrument background completely dominates the BF signal. 
On the contrary NESR are in the CCD corners and do not map the NXB spatial variations within the detector.
\cite{Moretti09}, using the data collected during the unique (and unrepeatable) observation performed with the camera shutter closed (September 2007), 
found a linear gradient in the NXB signal along the vertical direction of the CCD, with the bottom regions being 20$\%$  fainter than the top regions.
This gradient was found achromatic within the statistical  errors (1$\sigma\sim$5\%).
Here, we used all the BFs stacked data (6.3 Ms in total) to study the NXB spatial pattern. 
First, comparing the 4 NESR spectra we found that while the upper regions (NESR 1,2) and lower (NESR 3,4) are consistent 
between each other both in intensity  and spectral shape, the upper ones are significantly different from the lower ones (Fig.~\ref{fig:nxb_spec} ).
Then, to study the NXB spatial pattern we used the unresolved signal from all the BFs stacked data in the 7.0-10. keV energy band.
As said, in absence of bright sources, in this range, the signal is almost purely instrumental, dominated by the Nickel and Gold fluorescence lines (Fig.~\ref{fig:nxb_spec} ).
From these data we confirmed that the NXB has a linear vertical gradient with slope very well consistent with the one found by \cite{Moretti09}.
The scatter in the stacked BF 7-10. keV image  measured in 50$\times$50 pixel cells is $\sim$ 8\%, (1.5\% statistical); 
correcting the image by the linear gradient reduces this scatter to $\sim$ 2\% (Fig.~\ref{fig:nxbgrad} ).
If we assume that the NXB linear gradient is achromatic this means that, starting from the NESR signal, we can recover the 
NXB with $\sim$ 2.3\% accuracy (1 $\sigma$, plus the statistical uncertainty) in any position of the detector.
The assumption of achromaticity is justified by the analysis results of the shutter-closed observation.
%To estimate the expected NXB in our observations, we proceeded in the following way.
%First, we calculated the not-vignetted-exposure-map-weighted mean detector position of the spectral extraction region. Then the  expected NXB was 
%produced by  weighting the four  NESR spectra accordingly. 
%
%
%
%%%%%%%%%%%%%%%%%%%%%%%%%%%%%%%%%%%%%%%%%%%%%%%%%%%%%%%%%%%%%%%%%%%%%%%%%%%%%%%%%%
\subsection{Cosmic background}
\label{sec:cosmic_b}
\begin{figure}
\begin{tabular}{c}
\includegraphics[width=\columnwidth]{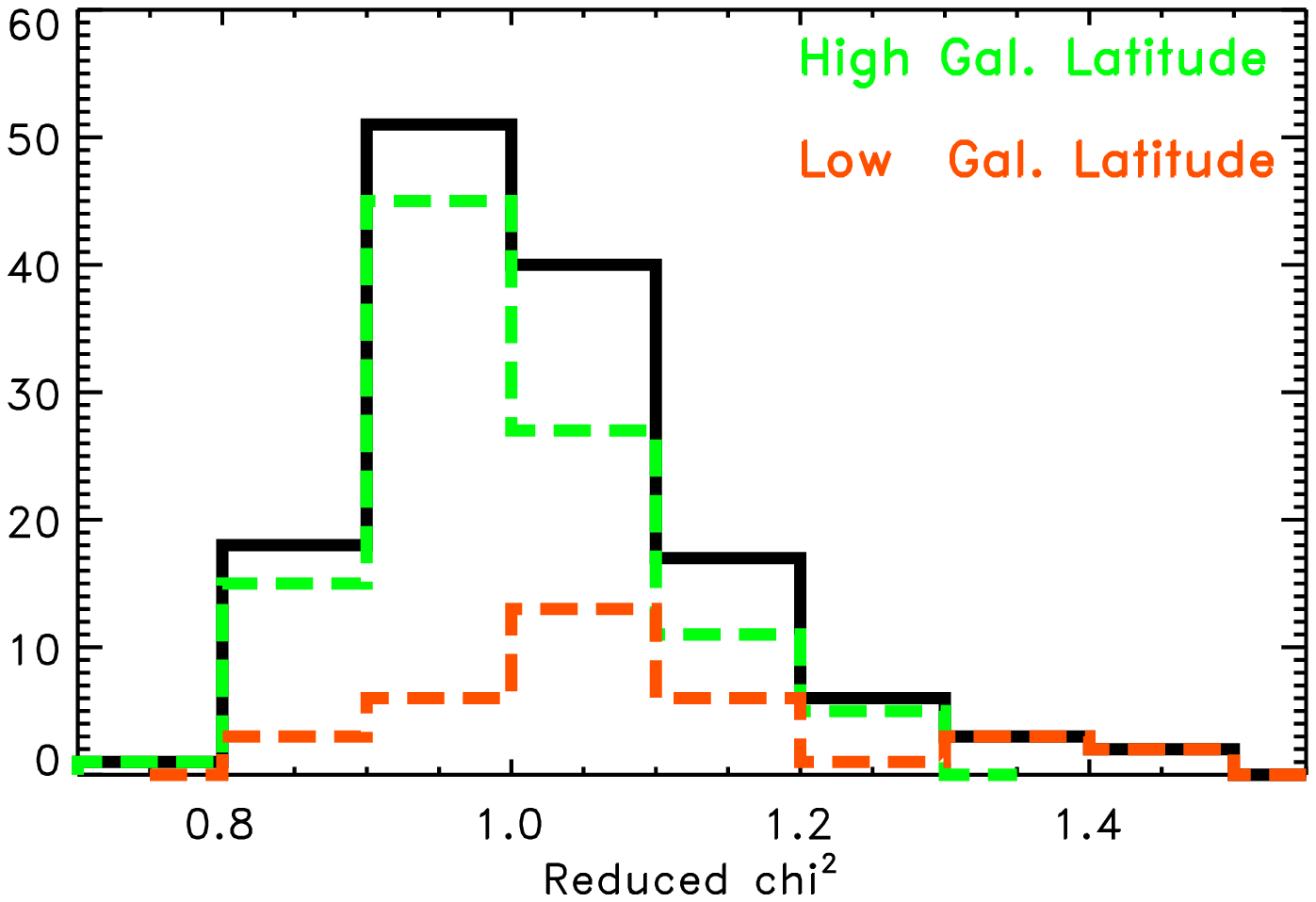} \\
\includegraphics[width=\columnwidth]{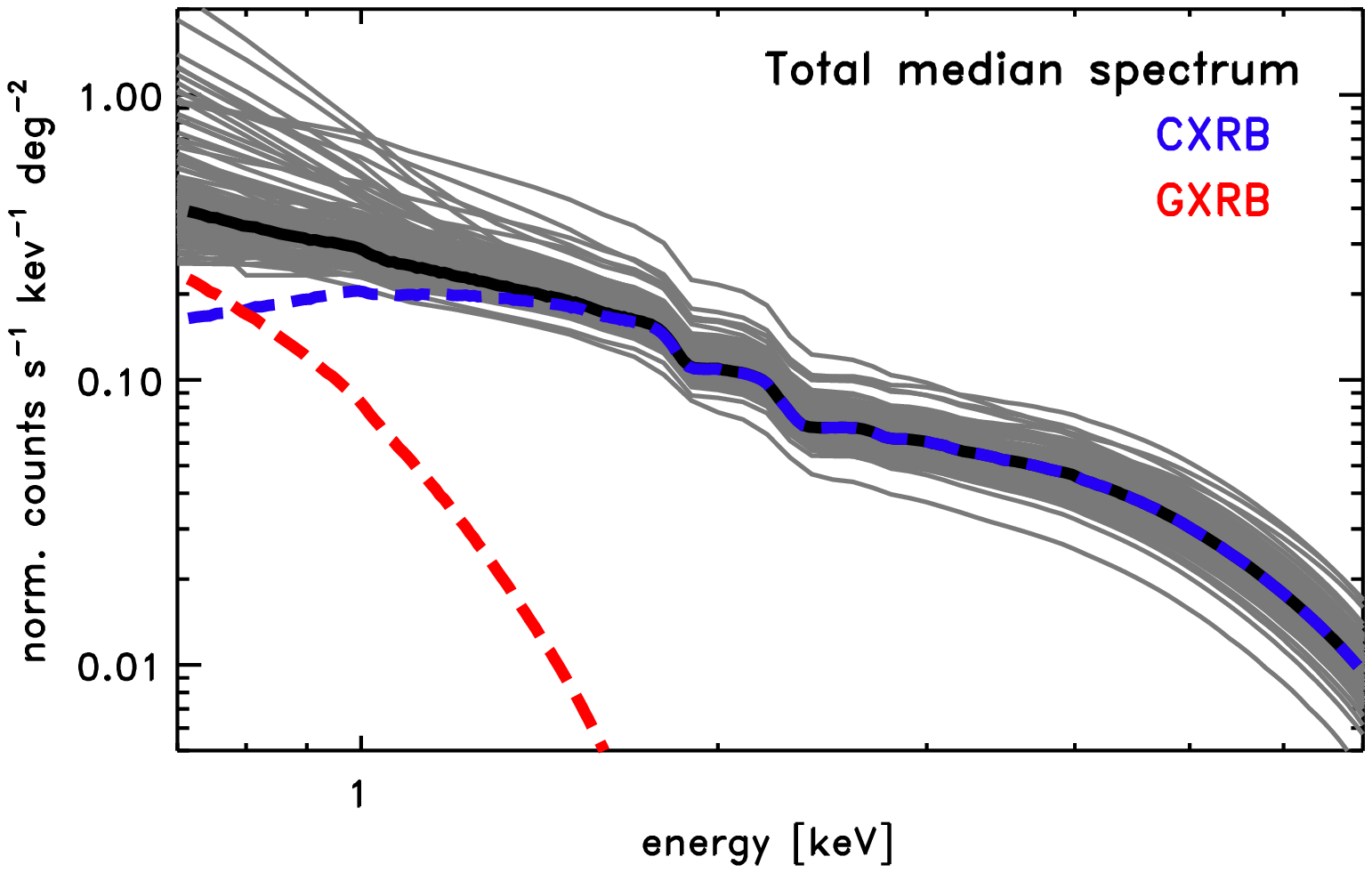} 
\end{tabular}
\caption{
{\bf Upper panel:}  The reduced $\chi^2$ distribution of the 135 fits, here split in high ($>$ 20 $^{\circ}$) and low Galactic latitude fields.
{\bf Lower panel:}  The 104 XRB models (grey lines) with the median spectrum in evidence (thick black line). For this we over plot the two components:
the Galactic bremsstrahlung (red dashed ) and the extragalactic power-law (blue dashed) }
\label{fig:fitmod}
\end{figure}
In order to study the statistical properties of the cosmic components of the background 
we performed the spectral analysis of the 135 BFs unresolved signal, using the NESR spectrum as instrumental background.  
Freezing the value of the Galactic absorbing Hydrogen column, for each field, to \cite{Kalberla05} values, we found that the unresolved emission in the 0.7-7.0  keV can always be well modeled  by a thermal component (bremsstrahlung) plus an absorbed  power law, accounting for GXRB and CXRB respectively.

\noindent 
From \cite{Snowden98} and \citep{Kuntz00}  we know that a physically motivated emission model of the XRB below 1 keV should consist of, at least, two 
components, the first of Galactic origin, the second from local hot bubble. To avoid this complexity in the modeling of the data, we limited 
our analysis to energies higher than 0.7 keV. 
Although not completely physically motivated, using a simple Bremsstrahlung provided us an accurate
phenomenological description of the data in this energy range with a limited number of parameters. This is suitable to our goal, whereas a physical description of 
the galactic thermal emission is beyond the scope of the present work (and the quality of our data at low energies).

\begin{figure}
\begin{tabular}{c}
\includegraphics[width=\columnwidth]{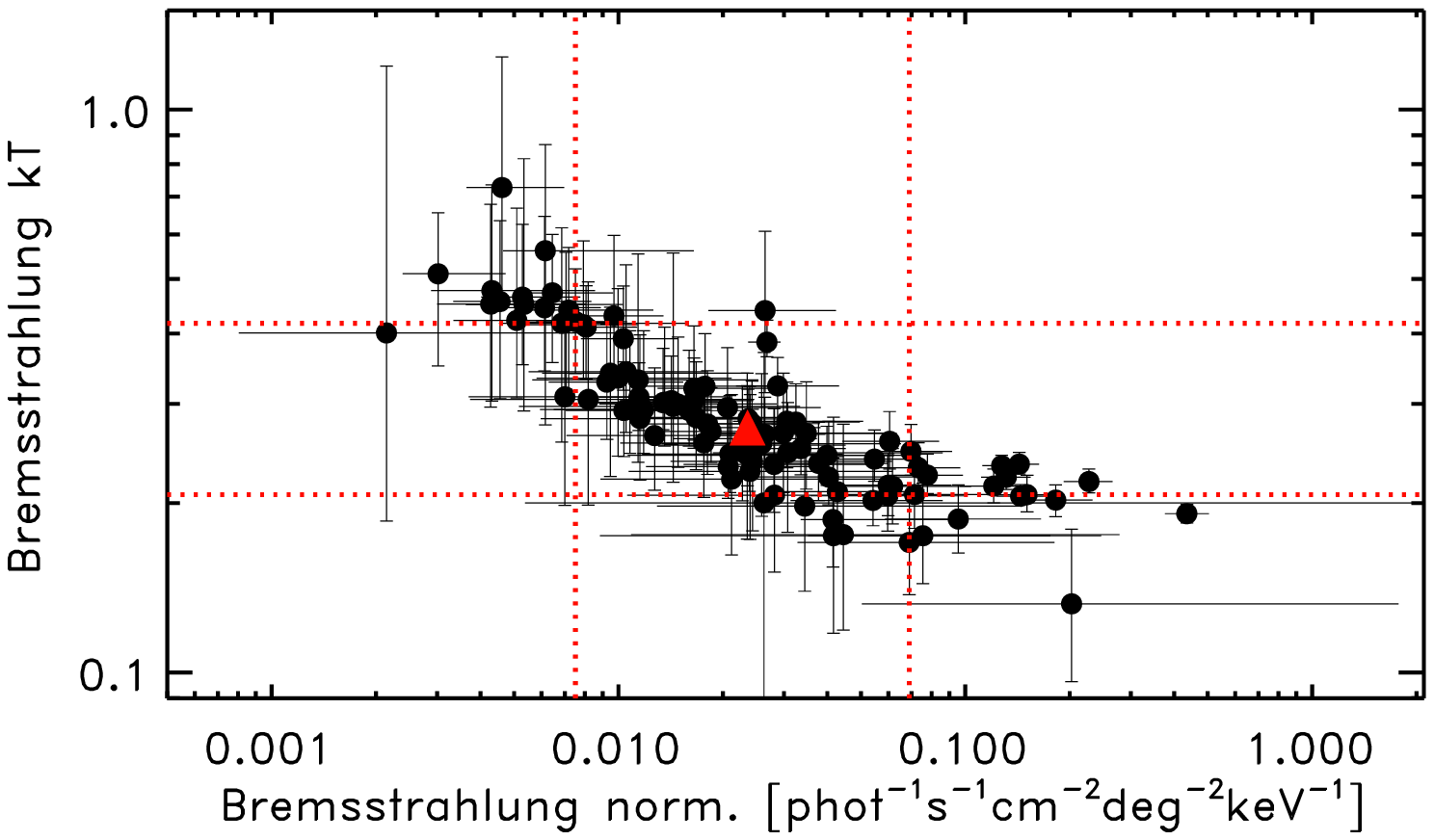} \\ 
\includegraphics[width=\columnwidth] {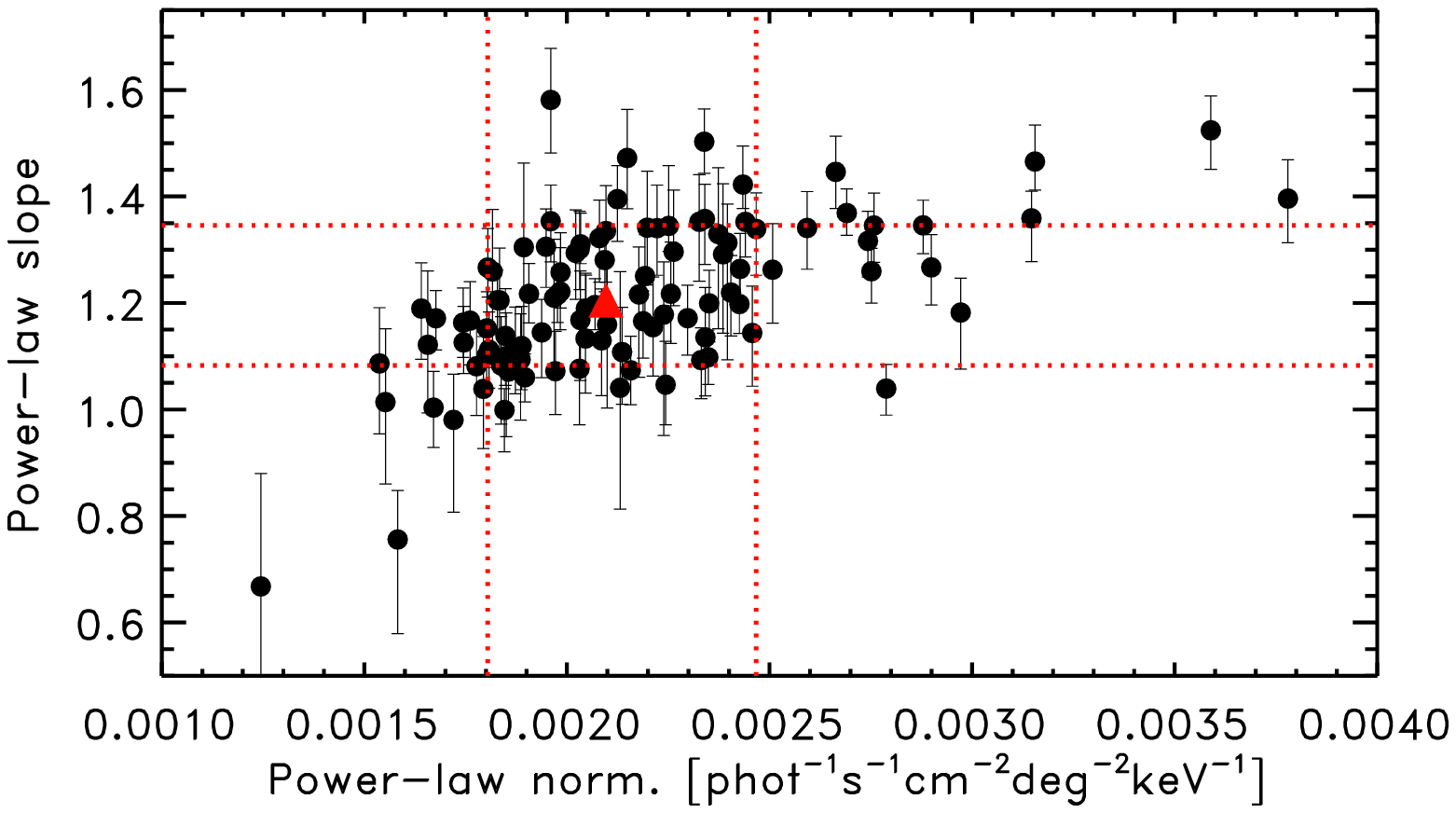} 
\end{tabular}
\caption{
{\bf Upper panel:}  The bremsstrahlung parameter space of the 104 high Galactic latitude fields. The red dotted lines represent the 16th and 84th percentile of the distributions and the red triangle the median. The red triangle represent the median values adopted as XRB model (Tab.~\ref{tab:medpha})
{\bf Lower panel:}  The power-law parameter space, with same notations. }
\label{fig:parfit}
\end{figure}

\noindent 
Indeed we found good  $\chi^2$ values for all the 135 fields (upper panel of Fig.~\ref{fig:fitmod}). Splitting  the sample in low ($<$ 20) and high latitude 
Galactic fields we found that fitting the latter always yielded   $\chi^2 ~<$1.3, while  the formers present slightly larger scatter. 
The soft thermal component typically contributes to $\sim$50\% of the total emission $\lesssim$1 keV, while it is negligible beyond 1.5 keV.
On the other hand, as we will see in the next Section, neglecting it, could significantly affect the temperature measurement in a regime where the 
cluster emission is comparable. As expected, we found that high and low latitude fields have different statistical properties, the latter sample 
presenting a larger scatter in the parameter distributions. 
As the cluster in the present sample are all (but PKS0745-19)  observed at high Galactic latitude we restricted our analysis to the extra Galactic fields.
The parameter space of the best fit values of the 104 high Galactic latitude BFs is shown in Fig.~\ref{fig:parfit}.
We used these data in two ways. First, from these distributions we derived the median spectrum (lower panel of Fig.~\ref{fig:fitmod}, 
Fig.~\ref{fig:parfit}  and Tab.~\ref{tab:medpha}) used as XRB background component in the cluster spectral analysis (see Section ~\ref{sect:spectra}). 
Then,  we used these models as input templates for the simulations employed to quantify the systematic temperature uncertainties 
(Section~\ref{sect:ssist}). 

\noindent 
Approximately  $\sim$20\% of the unresolved emission, in the XRT images, is due to stray light contamination from XRB sources outside the field of view \citep{Moretti09}. 
This contamination does not significantly affect our procedures, but it should be taken into account to compare the  normalization of our BF spectra  with the absolute 
measurements of the CXRB and GXRB. 
\begin{table}
\begin{center}
\caption[]{The median of the BF spectrum parameter distribution together with the 16th and 84th percentile values.}
\begin{tabular}{|l|ccc|}
\hline
                       & median & 16th \% & 84th \% \\
\hline
B. norm    [\small{phot s$^{-1}$cm$^{-2}$deg$^{-2}$kev$^{-1}$}]  &0.023&0.0075& 0.069 \\
B. kT   [keV]                                                                           &0.27&0.21& 0.42      \\
P.L. norm [\small{phot s$^{-1}$cm$^{-2}$deg$^{-2}$kev$^{-1}$}] &0.0021& 0.0018& 0.0024 \\
P.L. slope   []                                                                               &1.20&1.08&1.34     \\
\hline
\end{tabular}
\label{tab:medpha} 
\end{center}
\end{table}
\subsection{Time variability}
\label{sec:tvar}
\begin{figure}
\includegraphics[width=\columnwidth]{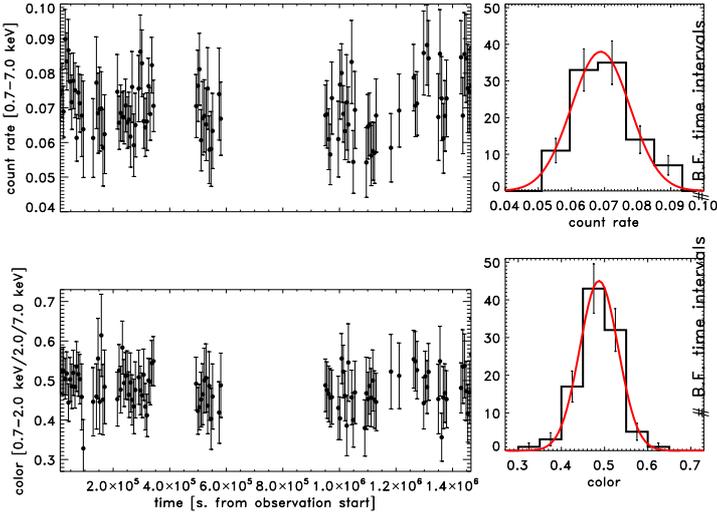}   
\caption{The light (upper-left panel) and color (lower-left) curve of the unresolved signal in the SWJ1557+35 observations. In the right panels we plot the corresponding distribution together with the gaussian fit. The observed scatter can be completely ascribed to the statistical error of the measurements. }

\label{fig:swjlc}
\end{figure}
Time variations of both cosmic and instrument background level are observed in the X-ray telescopes. In the Swift XRT particle flux variations in the satellite environment affecting the NXB,  are monitored and accounted for by the NESR data (see Sect~\ref{sect:nxb}). More insidious are the variations due to the solar wind exchange (SWCX) producing a diffuse photon emission at C, O, Mg, Ne  energies  with  different timescales from seconds to days. 
We quantified the impact of SWCX on Swift XRT observations studying the light curves in different energy bands both for the cluster data and for the BFs.

We studied the light curve in the soft band (0.7-2.0 keV) and the color curve  (0.7-2.0/0.7-7.0 keV) of the source free regions of the 2 serendipitous cluster observations (SWJ1557+35 and SWJ0847+13). We excluded the sources as we did for BFs and binned the data in  $\sim$ 1,500 s time intervals 
\footnote{Due to the low orbit of the satellite, the Swift XRT observations are composed by segments of  $\sim$ 1,500 seconds for each orbit (5700 s.).}.
These choices ensured  the necessary statistics to study the flux and spectral variations, as the signal registered in a BF  is $\sim 0.1$ count per second in the 0.7-7.0 keV energy band over the whole detector.
%We note that in this analysis we cannot subtract the NXB contribution due to the poor statistics of the signal in the NESR (0.003 counts per second).
Results relative to SWJ1557+35 are shown in Fig.~\ref{fig:swjlc}. The total exposure time of the observation is 200 ks which is split in 149 segments.
We considered only the 115 intervals with a duration longer than 1,000 s covering more than the 90\% of the total duration (185 ks). We found that both 
the flux and the color scatter are well consistent with the statistical one.
In other words, in this data set, we did not find any  detectable signature of SWCX. 
For SWJ0847+13 we found similar results, while for  the four remaining objects the results of this analysis are not conclusive,  
due to the very high level of the  cluster signal over the entire field of view. 

\begin{figure}
\includegraphics[width=\columnwidth]{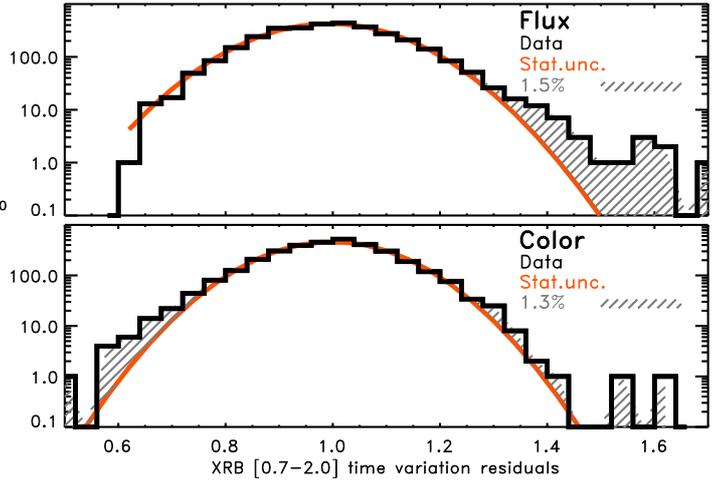}   
\caption{The residual distribution of the flux and color XRB curve of a collection of BFs. We put in evidence the expected Gaussian
distribution  from statistical uncertainties and the departures from this. We found that only $\sim$ 1.5\% of all the time intervals considered can be considered outliers.  }
\label{fig:timesist}
\end{figure}

We extended this kind of analysis to all the BFs. 
To get rid of the cosmic variance, we normalized each observation to its median value
and compared the residuals. We found that only $\sim$ 1.5\% of all the time intervals considered can be identified outliers Fig.~\ref{fig:timesist}.
Given these findings, we decided to neglect the effect of the  SWCX in both our spectral analysis and in our simulations. 
%
%%%%%%%%%%%%%%%%%%%%%%%%%%%%%%%%%%%%%%%%%%%%%%%%%%%%%%%%%%%%%%%%%%%%%%%%%%%%%%%%%%%%%%%%%%%%%%%%%%%%%%%%%%%%%%%%%%%%%%
\section{Systematic uncertainties in temperature measurement}
\label{sect:ssist} 
In order to evaluate the systematics in the temperature measurement we 
used \texttt{XSPEC}(v12.5) to simulate a set of cluster+XRB \texttt{bremss}+\texttt{wabs}$\times$(\texttt{pow}+\texttt{apec}) 
spectra, as observed by the Swift XRT.

\noindent
To simulate the XRB we used BFs, but we had to consider two further complications.
First, CXRB normalization and spectral slope are expected to vary according the image flux limit: the spectrum of the unresolved CXRB emission in deeper images 
is expected to be fainter and harder \citep{Moretti03}. Second, the CXRB variance depends on the size of the extraction regions \citep{Kushino02, Revnivtsev08, Moretti09}.  
To account for these effects we split the BF sample in five exposure time bins (10-30, 30-50, 50-75, 75-125, 125-200 ks) and 
for each BF we extracted the XRB spectrum from  regions of different sizes (1.0, 0.5, 0.25 and 0.1 times the entire field of view). 

\noindent
To simulate the cluster emission we used 37 different cluster temperatures (in the range 1-10 keV, step 0.25 keV)  with the same flux (8$\times$10$^{-13}$ erg s$^{-1}$ cm$^{-2}$ )
and exposure time (100 ks) in order to collect  a number of source photons to  make the statistical errors negligible ($>$ 10000 source counts ). 
Then, for each spectrum we modified the source/background ratio, varying the BACKSCAL keyword, in order to simulate different values of surface brightnesses:
10 steps  between  2$\times$10$^{-12}$ and 8$\times$10$^{-11}$ erg s$^{-1}$ cm$^{-2}$ deg$^{-2}$).
For each step of this 37$\times$10 grid we simulated 200 different realizations letting the metallicity and N$\rm_{H}$ normally and randomly varying around
the mean values of 0.3 Z$_{\odot}$ and 3$\times$10$^{20}$cm$^{-2}$ respectively with a scatter of 0.06 (20\%) and 6$\times$10$^{19}$(20\%).
Thus, for each simulated cluster we summed an XRB model randomly choosing one of the BF model cycling over the
exposures $\times$ region size grid: this resulted in a total of 200 spectra $\times$ (37$\times$10) clusters ($\times$ 5$\times$4) XRB grid.
Finally, to each simulation, we added the instrumental background stacking a collection of the NESR data, randomly re-normalizing it by a gaussian
deviation of 2.3\% to account for the uncertainties in the reproduction of the spatial pattern (Section~\ref{sect:nxb}).
We used all the BF observed in 2009, for a total of $\sim$1.5 Msec, in order to maximize the statistics and, at the same time, to avoid 
particle background variation on $\sim$  1 year time scale , as observed in \citet{Moretti09}, which are probably due to variation in the Solar activity.

\noindent
We fit the simulated data using the same procedure we used for our real data as described in Sect~\ref{sect:spectra}. 
We used a \texttt{bremss}+\texttt{wabs}$\times$(\texttt{pow}+\texttt{apec}) model, freezing the XRB parameters to the  
BF  XRB median spectrum (Sect. \ref{sec:cosmic_b} ), just normalized for the input area. 
Moreover, we froze the N$\rm_{H}$ and metallicity to the mean values (0.3 Z$_{\odot}$ and 3$\times$10$^{20}$cm$^{-2}$ respectively ) and we used 
the GRB 090618 follow-up  observation NXB as background. 

\begin{figure}
\includegraphics[width=\columnwidth]{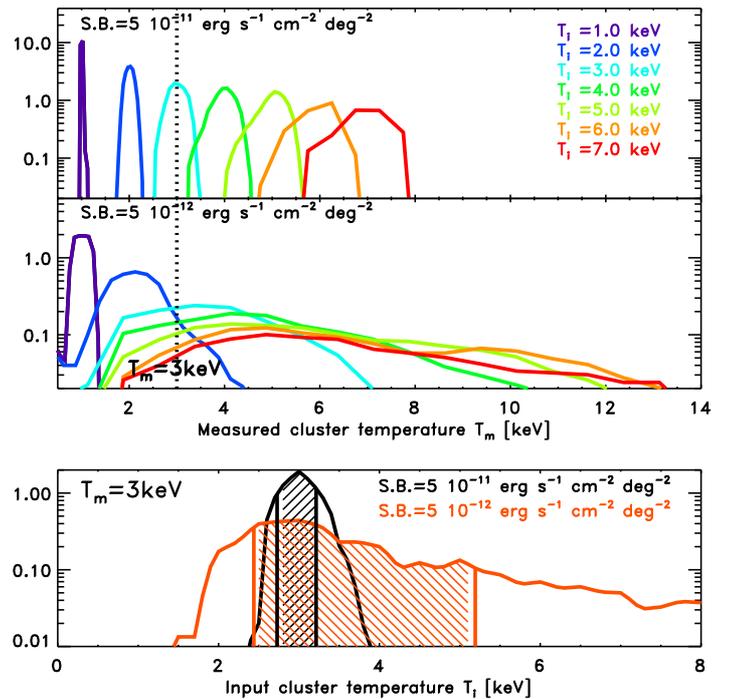}   
\caption{{\bf Upper panel:} The simulation outputs relative to an APEC spectrum with kT=1-7 keV 100 ks observation extracted from a
region size of 0.5 the field of view in a high signal to background ratio regime (SB = 5 10$^{-11}$ erg s$^{-1}$ cm$^{-2}$ deg$^{-2}$).
Different colors indicate the distribution of the results of the fits of different input temperatures (200 realizations each).
{\bf Middle panel:} The same of the upper panel in the case of low signal to background ratio 
(SB = 5 10$^{-12}$ erg s$^{-1}$ cm$^{-2}$ deg$^{-2}$). 
{\bf Lower panel:} The distribution of the T$_i$ scatter values measured at T$_m$=3 keV relative to the two different values of 
surface brightness; the intervals between 16th and 84th percentile of both distributions are in evidence. }
\label{fig:tsist}
\end{figure}

\noindent
As an example of the simulation outputs, in Fig. \ref{fig:tsist}, we plot the distribution of the results of the fit on the 200 realizations 
of APEC spectra with different input temperatures T$_i$, for two different values of surface brightness (high surface brightness in the higher panel,
low surface brightness in the middle panel).
In this case the background was extracted from BF with exposure time within 75-125 ks
and size $\sim$ 0.5 the field of view. It is immediately evident that at lower values of SB the scatter of the 
simulation outputs is significantly higher. 
\noindent
We used these outputs to calculate the systematic uncertainties in our cluster temperature profiles in the following way. 
We did not calculate the error of a measurement T$_m$ simply as the scatter of the results of the fit of the 200 realizations at 
temperature T$_m$. Instead we calculated it from the distribution of the "true" temperatures T$_i$ that have a non null probability 
of being measured as T$_m$. 
When the SB is high ($>$ 1e-11 erg s$^{-1}$ cm$^{-2}$ deg$^{-2}$) the systematic scatter of the measurement T$_i$ is very
small and the probability that a true temperature T$_i \gg$ T$_m$ would yield a T$_m$ measurement is very low.
For example in the case shown in the upper panel of Fig.~\ref{fig:tsist} with a SB $\sim$ 5 10$^{-11}$erg s$^{-1}$ cm$^{-2}$ deg$^{-2}$ 
the probability that a cluster region with a true temperature T$_i$ $>$4 would be measured 3 keV is almost null. 
On the contrary, when the SB is low and the background is comparable or higher than the cluster signal, 
the scatter of the measurement can be very large, especially at high temperatures. 
In the middle panel of the same figure, we show the same example for a SB which is a factor 10 lower (5 10$^{-11}$erg s$^{-1}$ cm$^{-2}$ deg$^{-2}$):
in this case the probability that T$_i$ $>$4 would be measured 3 keV is not negligible. 
Thus, to calculate the T$_m$ error, for each T$_i$ of the simulation grid, we calculated the T$_i$ output distribution value in T$_m$; 
then we calculated the errors as the 16th and 84th percentile of this distribution (lower panel of Fig.\ref{fig:tsist}).

\noindent The systematic errors reported in Tab.~\ref{tab:dataprof} and plotted in Fig.~\ref{fig:tprof} are calculated in this way.
In most of the spectra of the present sample, the systematic errors do not strongly affect the total error budget of our measurements 
as the surface brightness  of the observed regions is higher than 10$^{-11}$ erg s$^{-1}$ cm$^{-2}$ deg$^{-2}$.
The external regions of the two more distant clusters are the only cases where SB is lower; but here the statistical errors are very large too. 
However the analysis of the temperature measurement systematic errors is mandatory to realistically 
calculate the expected errors in the outer regions assuming to have a good statistics data-set as in our simulation of Abell1795.   

\noindent
\cite{Leccardi07} shown that, using  the most common likelihood estimators, the measurements of the temperatures from outer regions of
galaxy clusters are strongly biased.  We note that our way of calculating the temperature measurement uncertainties represents a refinement 
of the usual procedure used in the literature.
Indeed, the current approach is a particular case of our procedure, given by the simplifying assumption that  the measured value T$_m$ coincide 
with the true value T$_i$. In the remaining we will refer to this way of calculating the error as H1, while we will call our refined procedure 
H2.
%%%%%%%%%%%%%%%%%%%%%%%%%%%%%%%%%%%%%%%%%%%%%%%%%%%%%%%%%%%%%%%%%%%%%%%%%%%%%%%%%%%%%%%%%%%%%%%%%%%%%%%%%%%%%%%%%%%%%%%
\section{A1795: a case study}
\label{sect:tsimu} 
Chandra and XMM-Newton deep observations showed that Abell 1795 is a dynamically relaxed system within r$_{500}$ \citep{Vikhlinin06}, 
although a 50 kpc (40 \arcsec) long X-ray filament in the core is present \citep{Fabian01}.  
Thanks to the lower background, Suzaku, for the first time, succeeded in extending the observations in the radial range  r$_{500}$ and  r$_{200}$
finding a significant asymmetry between the northern and southern part of the cluster. Departure form the hydrostatic equilibrium, with the plasma 
in-falling from the northern skirt is a possible explanation \citep{Bautz09}. Suzaku observation sensitivity is limited by two factors: the  background 
from solar wind  (SWCX) and the XRB variance \citep{Bautz09}. The latter was partially mitigated exploiting XMM-Newton or Chandra observations.

\noindent In the previous Section we described the procedure we developed to quantify the impact of these two factors on Swift XRT 
temperature measurements for a sample of cluster observed in their central regions. Here we use the same procedures to calculate 
how accurate a temperature measurement performed 
by a fairly deep (300 ks) Swift XRT observation of the northern (the brighter) out-skirt of Abell 1795 would be. 
\begin{table*}
\begin{center}
\caption[]{Extraction regions and spectral analysis results of the 300 ks simulated observation. For each annulus we report the radii,
the area, the effective exposure time (accounting for exposure maps and vignetting), the expected mean cluster surface brightness,
 the expected total events and  the source/total ratio and the expected errors. 
We report explicitly the relative contribution of the statistics and the systematics terms.
 The latter are calculated either in the assumption that Suzaku measurement is the true value 
(H1) either considering the distribution of the possible  true temperature values as expected from our simulation (H2) as explained in Sect. \ref{sect:ssist}. 
The errors are reported at 90\% of confidence consistently with Suzaku published numbers \citep{Bautz09} }

\begin{tabular}{|ll|ccccccccc|}
\hline
  ann. rad.   & ann. rad.     &  area         &  exp.  &  Surf. Bright.                                &  Tot. evt &   S/(S+B) &  T stat. error       & T syst. error (H1)   & T syst. error (H2)    \\
                  & [r$_{200}$]  & [deg$^2$]  &  [ks]   & erg s$^{-1}$ cm$^{-2}$ deg$^{-2}$ &              &       [\%]       &   [frac.]         & [frac.]                   & [frac.]    \\          
 \hline
12.6-13.3 & 0.49-0.52 & 1.1E-02 & 8.7E+01 & 2.4E-11 & 1504.4 & 69.72 & 0.09-0.10 & 0.15-0.22 & 0.15-0.15 \\
13.3-14.0 & 0.52-0.54 & 1.3E-02 & 8.7E+01 & 2.0E-11 & 1526.7 & 66.05 & 0.09-0.11 & 0.15-0.22 & 0.13-0.15 \\
14.0-14.9 & 0.54-0.58 & 1.6E-02 & 8.6E+01 & 1.6E-11 & 1631.3 & 61.37 & 0.08-0.10 & 0.19-0.32 & 0.13-0.55 \\
14.9-16.0 & 0.58-0.62 & 1.9E-02 & 9.0E+01 & 1.3E-11 & 1800.0 & 55.64 & 0.09-0.10 & 0.22-0.41 & 0.17-0.56 \\
16.0-17.4 & 0.62-0.68 & 1.9E-02 & 1.1E+02 & 1.0E-11 & 1909.6 & 49.41 & 0.12-0.13 & 0.20-0.34 & 0.14-0.59 \\
17.4-19.0 & 0.68-0.74 & 1.9E-02 & 1.2E+02 & 1.1E-11 & 2197.0 & 52.07 & 0.07-0.06 & 0.20-0.36 & 0.12-0.60 \\
19.0-20.6 & 0.74-0.80 & 1.5E-02 & 1.5E+02 & 9.2E-12 & 2070.5 & 47.39 & 0.07-0.11 & 0.21-0.37 & 0.09-0.61 \\
20.6-22.4 & 0.80-0.87 & 1.3E-02 & 1.8E+02 & 9.0E-12 & 2080.4 & 46.78 & 0.07-0.12 & 0.21-0.37 & 0.10-0.60 \\
22.4-25.0 & 0.87-0.97 & 1.7E-02 & 1.8E+02 & 6.4E-12 & 2339.5 & 38.39 & 0.09-0.10 & 0.23-0.39 & 0.20-1.03 \\
\hline
\end{tabular}
\label{tab:simprof} 
\end{center}
\end{table*}

We used the (northern) surface brightness and temperature radial profile as measured by Suzaku up
to 25$\arcmin$ and reported by \citep{Bautz09} to simulate the cluster surface brightness profile for 
a 300 ks Swift XRT exposure.  This was done using a real exposure map  in order to account for vignetting, 
CCD defects and different aim points and roll angles of a real observation.
Because the Abell 1795 virial radius largely exceeds the field of view (r$_{200}$=25.7$\arcmin$), a combination of different pointing 
is necessary for our purposes.  We found that a mosaic of two different aim points with displacement of 13 $\arcmin$ from the cluster center optimizes the observation
efficiency, providing  us with the maximum exposure in the 20$\arcmin$-25$\arcmin$ annulus. 
Starting from r$_{200}$,  we calculated the annuli which would contain 1000 events in the 0.5-2.0 energy band in order to keep the statistical error
at the level of  $\lesssim$10\%

\noindent We used, as XRB, the same of \cite{Bautz09}  which was measured in the most external observed region and is very close 
and consistent with our median XRB; as NXB the NESR signal registered during the observation of GRB 090618 chosen as the 
most recent observation lasting more than 300 ks,

\noindent  
Thus, for each annulus we calculate the total (statistical + systematics) error in temperature measurement using the recipes described in Sect.~\ref{sect:ssist}.

\begin{figure}
\includegraphics[width=\columnwidth] {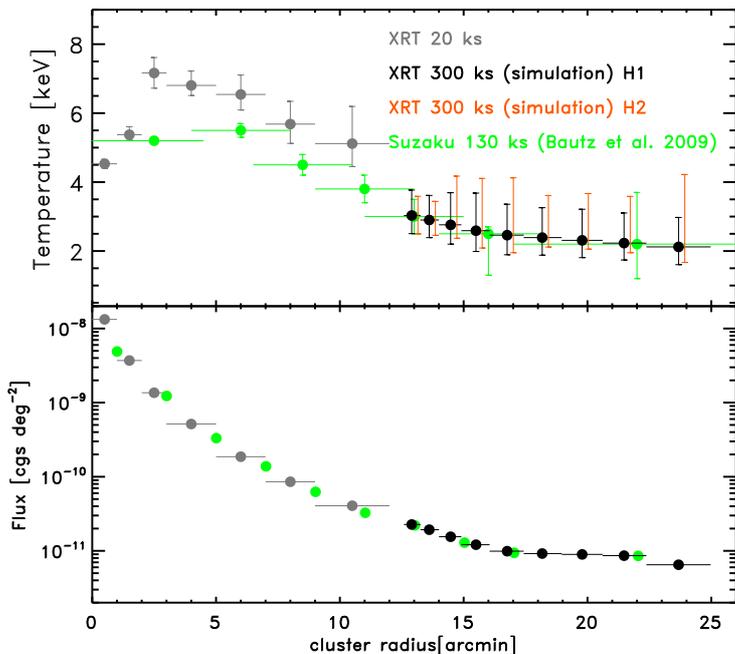}     
\caption{Simulation of 300 ks observation of the northern skirt of Abell 1795. {\bf Upper panel:} Surface brightness of the cluster. Green points are from 
Suzaku observation (\cite{Bautz09}). Grey points are from Swift XRT observation (already shown in Fig.~\ref{fig:tprof}). Black points
would be the result of 300 ks observation. 
{\bf Lower panel:}  The surface brightness profile of the cluster as observed by Suzaku, with the same color code. XRT spatial bins in the 
simulation are chosen in order to have a minimum of 1000 source counts in the 0.7-2.0 keV band.}
\label{fig:a1795_prof}
\end{figure}

\begin{figure}
\includegraphics[width=\columnwidth] {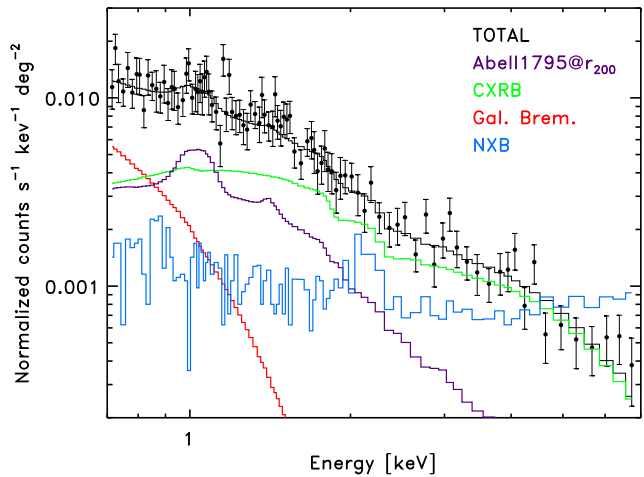}     
\caption{Simulation of the 300 ks of the last annulus.}
\label{fig:a1795_spec}
\end{figure}

\noindent Results of this procedure are reported in Tab.~\ref{tab:simprof} and shown in the upper panel of Fig.~\ref{fig:a1795_prof} in comparison with the Suzaku 
measurement. Systematic errors were calculated, first, assuming the Suzaku measurement as the true value of the temperature. This  was done  following 
the approach of \cite{Bautz09} to compare the Swift XRT expected accuracy with Suzaku (H1 column in Tab.~\ref{tab:simprof} and black error bars in Fig.~\ref{fig:a1795_prof} ).  
The simulated spectrum from the external bin is shown in Fig.~\ref{fig:a1795_spec}.
We found that a 300 ks Swift-XRT observation would significantly improve the accuracy of the Suzaku temperature measurement 
in the (northern) outskirts of the cluster Abell 1795, both in terms of spatial binning and relative accuracy.
Indeed in the annulus within 17-25 $\arcmin$ while Suzaku could measure only one single temperature with  an accuracy of $\sim$ 60\%, we expect 
that the XRT observation would be able to measure four different temperatures with an accuracy of $\sim$ 40\%.

\noindent Second, we refined the error calculation, also considering the distribution of the possible true temperatures 
which could yield the measured value (H2 procedure, see Sect.~\ref{sect:ssist}). In this case the upper error bars are significantly larger. This is the result 
of the not negligible bias in the high temperature (T $>$ 5 keV) measurement in the low SB regime (Sect. \ref{sect:ssist} and Fig~\ref{fig:tsist}).
In this case in the three bins within 17-23 $\arcmin$ we expect an accuracy of $\sim$ 60\% similarly to the Suzaku one, while in the last bin 
(23-25 $\arcmin$) the upper error bar is $\sim$ 100\% of the measured value (H2 column in Tab.~\ref{tab:simprof} and red error bars in Fig.~\ref{fig:a1795_prof} ).
We note that the impact  of the bias in the high temperature (T $>$ 5 keV) measurement in a low SB regime on temperature accuracy, 
has never been quantified for Suzaku. Even in the unrealistic assumption that this is completely negligible, the Swift XRT observation would improve the 
accuracy of the temperature profile significantly narrowing the spatial  binning up to $\sim$ R$_{\rm200}$.

\noindent
 In a spatially resolved spectral study of a low surface brightness source, a telescope reaches the limit 
of its capabilities, when a deep observation allows to map the source at the best of its angular resolution 
(bins $\geq$3 times the HEW, to avoid PSF mixing)  and, at the same time, in each single bin, the collected photons are  
enough to make the statistic error negligible with respect to the systematics (mainly the background ones).
The Suzaku telescope, in its observation of Abell 1795, almost reached its limits: in fact, most of the error in the temperature
measurement, is due to background variance, while the angular resolution can be only slightly improved as the telescope 
HEW is 2\arcmin. 
On the other hand, we showed that, with a 300 ks observation, Swift-XRT  could reach the same level of accuracy 
with an angular resolution of $\sim$ 2\arcmin. This leaves a wide range of improvement: indeed, doubling the 
observation exposure would allow to halve the spatial binning, as the HEW is 18\arcsec. 
%
%%%%%%%%%%%%%%%%%%%%%%%%%%%%%%%%%%%%%%%%%%%%%%%%%%%%%%%%%%%%%%%%%%%%%%%%%%%%%%%%%%%%%%%%%%%%%%%%%%%%%%%%%%%%%%%%%%%%%%%
\section{Discussion and conclusion}
\noindent
In the outer regions of nearby clusters, the ICM emission is only a small fraction of the whole signal collected by the detector. 
The regime, where the background systematics, affect the spectroscopic measurements much more than the statistical error, is 
easily reached. 
While XMM-Newton and Chandra are not suitable for this kind of observation due to the high level of particle background, 
in all the published works presenting Suzaku observations of cluster outskirts the evaluation of the XRB and its variance
is the main issue. Different approaches have been pursued. \cite{Bautz09} and \cite{Hoshino10},  in their studies of Abell 1795 and Abell 1413 
respectively, used the signal from external regions at $\sim$ 1.2 $\times$r$_{200}$ as XRB, in the assumption that the cluster emission is 
negligible at that distance from the center. 
To study the temperature profile of PKS0745-19, \cite{George09}  used the Lockman Hole observation which was performed just few days before the 
cluster observation. Interestingly they found significant emission from the cluster ICM at distance $>$ 1.5 $\times$r$_{200}$.
\cite{Kawaharada10} used the two closest observations (at $\sim$ 8$^\circ$ of distance) among the ones suitable as blank fields in the Suzaku archive.
\cite{Reiprich09} and \cite{Fujita08} used the classical models from literature to fit the cluster spectrum together with the XRB (using a
different choice of free parameters). 

\noindent
In this paper we presented for the first time the analysis of the Swift-XRT observations of a sample of 6 galaxy clusters. 
We measured the temperature profiles as far as $\sim$ 0.5 R$_{200}$ for all (but Coma)  sample clusters. 
To estimate the cosmic background we used a statistical approach, modeling a representative sample of blank fields,
with a sky-coverage of $\sim$ 15 deg$^2$. 
We used the blank field median spectrum as XRB model to fit our cluster spectra. 
We calculated the systematics of this approach by simulating realistic clusters with different 
temperatures and surface brightnesses summed to real XRB data extracted from blank fields with different sizes and exposure times. 
With this approach in the systematics calculation,  exploiting a statistically fair sample of BF,  we directly accounted 
for the XRB  variance  (for both CXRB and GXRB ).
Moreover we presented a new way to calculate the uncertainties in the temperature measurements significantly refining the current approach in
literature. Given a measure T$_m$, we accounted for the possibility that T$_m$ is produced by a T$_i \neq$ T$_m$.
This  allowed us to realistically simulate the temperature measurement in the outer regions of  Abell 1795 
which would be provided by a deep XRT observation.  We showed that, thanks to an unprecedented combination of low 
background, good PSF the Swift XRT would be able to significantly improve the current accuracy of the temperature 
measurements in the outer regions of nearby clusters.   

\noindent
The ideal telescope for  cluster outskirts observation would be a large grasp (wide field and large collecting area) 
and low background telescope such the proposed WFXT \citep{Murray10}. 
In the next decade eRosita will be the only mission operative with these characteristics \citep{Predehl10}, a grasp 10 times 
larger than XMM (100 times larger than Swift-XRT). Interestingly, eRosita, with an effective area of $\sim$ 1500 cm$^2$ 
at 1.5 keV ($\sim$10 times larger than XRT) and an expected background of $\sim$ 9 counts s$^{-1}$ deg$^{-2}$ ($\sim$10 times larger than XRT)
will have the same source / background ratio of the Swift-XRT when observing extended  sources. 
If these numbers will be confirmed in flight, and the NXB will be reproduced with the same accuracy ($\lesssim$ 3\%),
at a given value of surface brightness, eRosita will have systematic errors on temperature measurements 
which will be very close to the ones we found for Swift-XRT.
In this case, our proposed XRT observations would represent a pilot for the eRosita mission; 
on the other hand, if the NXB of eRosita will be higher than expected (eRosita will be the first X-ray telescope positioned in L2) 
or it will be impossible to reproduce it with the same accuracy, the XRT observation would remain the only way to improve 
our knowledge of the cluster outskirts physics at least for the next decade. 
%In this regime even a small telescope as the Swift XRT easily reaches a regime where the  background systematics affect 
%the spectroscopic measurements much more than the statistical error.
%Indeed, in spite of the much smaller area, the Swift XRT is more suitable for this kind of observation with respect to Chandra and XMM-Newton.  
%In fact the XRT background is such that at the virial radius of nearby clusters the ICM emission would be $\sim$30\% of the total signal.
%As pointed out by \cite{Bautz09} the main limitation of Suzaku is the poor PSF. First, the poor PSF, even in a very high statistic regime,  
%allows only a coarse probing ($>$4$\arcmin$) of the ICM. 
%It also prevents from removing point-like sources present along the line of sight of the clusters adding an highly 
%significant term in the background systematic uncertainties. On the contrary, 
%the Swift XRT good PSF would allow a finer spatial sampling and to significantly reduce the expected level of the CXRB and its variance,
%if the smaller effective area is compensated by longer exposures.
%Moreover, thanks to the  NXB background monitoring provided by the NESR, the Swift-XRT allows a good control also on the instrumental 
%fraction of the background systematics ($\sim$ 2\%). 
\noindent %
\begin{acknowledgements}
\noindent We acknowledge the financial contribution from contracts ASI-INAF I/023/05/0,  I/088/06/0,  I/011/07/0 "

\noindent This research has made use of 

\noindent -- the X-Rays Clusters Database (BAX)
which is operated by the Laboratoire d'Astrophysique de Tarbes-Toulouse (LATT),
under contract with the Centre National d'Etudes Spatiales (CNES);

\noindent -- the NASA's Astrophysics Data System;

\noindent -- the NASA/IPAC Extragalactic Database (NED) which is operated by the Jet Propulsion Laboratory, California Institute of Technology, under contract with the National Aeronautics and Space Administration.

\end{acknowledgements}
%%%%%%%%%%%%%%%%%%%%%%%%%%%%%%%%%%%%%%%%%%%%%%%%%%%%%%%%%%%%%
%%%%% References %%%%%
\bibliographystyle{aa}   %>>>> makes bibtex use spiebib.bst
\bibliography{tot09}   %>>>> bibliography data in report.bib
%\clearpage
\end{document}